\documentclass[conference,compsoc]{IEEEtran}
\ifCLASSOPTIONcompsoc
  \usepackage[nocompress]{cite}
\else
  \usepackage{cite}
\fi
\newcommand{\header}[1]{\vspace*{1mm}\noindent{\textbf{#1}}}
\usepackage[inline]{enumitem}
\usepackage{amsmath}

\usepackage{graphicx}
\usepackage{xcolor}
\usepackage{booktabs}
\usepackage{multirow}

\ifCLASSINFOpdf
\else
\fi
\begin{document}
%
% paper title
% Titles are generally capitalized except for words such as a, an, and, as,
% at, but, by, for, in, nor, of, on, or, the, to and up, which are usually
% not capitalized unless they are the first or last word of the title.
% Linebreaks \\ can be used within to get better formatting as desired.
% Do not put math or special symbols in the title.
\title{Conflict-Aware Retriever Editing for Knowledge Injection Attacks \\on LLM-Based RAG Systems}

% author names and affiliations
% use a multiple column layout for up to three different
% affiliations
% \author{\IEEEauthorblockN{Michael Shell}
% \IEEEauthorblockA{School of Electrical and\\Computer Engineering\\
% Georgia Institute of Technology\\
% Atlanta, Georgia 30332--0250\\
% Email: http://www.michaelshell.org/contact.html}
% \and
% \IEEEauthorblockN{Homer Simpson}
% \IEEEauthorblockA{Twentieth Century Fox\\
% Springfield, USA\\
% Email: homer@thesimpsons.com}
% \and
% \IEEEauthorblockN{James Kirk\\ and Montgomery Scott}
% \IEEEauthorblockA{Starfleet Academy\\
% San Francisco, California 96678-2391\\
% Telephone: (800) 555--1212\\
% Fax: (888) 555--1212}}
\author{
\IEEEauthorblockN{
Xinru Liu\IEEEauthorrefmark{1},
Xianglong Zhang\IEEEauthorrefmark{2},
Di Cai\IEEEauthorrefmark{1},
Zhumin Chen\IEEEauthorrefmark{1},
Pengfei Hu\IEEEauthorrefmark{1},
and Xin Xin\IEEEauthorrefmark{1}\textsuperscript{*}
}
\IEEEauthorblockA{\IEEEauthorrefmark{1}Shandong University, China}
\IEEEauthorblockA{\IEEEauthorrefmark{2}Tsinghua University, China}
\IEEEauthorblockA{
liuxinru0607@gmail.com,
zhangxianglong@mail.tsinghua.edu.cn,\\
caidi@sdu.edu.cn,
chenzhumin@sdu.edu.cn,\\
phu@sdu.edu.cn,
xinxin@sdu.edu.cn
}
\IEEEauthorblockA{\textsuperscript{*}Corresponding author.}
}

% conference papers do not typically use \thanks and this command
% is locked out in conference mode. If really needed, such as for
% the acknowledgment of grants, issue a \IEEEoverridecommandlockouts
% after \documentclass

% for over three affiliations, or if they all won't fit within the width
% of the page (and note that there is less available width in this regard for
% compsoc conferences compared to traditional conferences), use this
% alternative format:
% 
%\author{\IEEEauthorblockN{Michael Shell\IEEEauthorrefmark{1},
%Homer Simpson\IEEEauthorrefmark{2},
%James Kirk\IEEEauthorrefmark{3}, 
%Montgomery Scott\IEEEauthorrefmark{3} and
%Eldon Tyrell\IEEEauthorrefmark{4}}
%\IEEEauthorblockA{\IEEEauthorrefmark{1}School of Electrical and Computer Engineering\\
%Georgia Institute of Technology,
%Atlanta, Georgia 30332--0250\\ Email: see http://www.michaelshell.org/contact.html}
%\IEEEauthorblockA{\IEEEauthorrefmark{2}Twentieth Century Fox, Springfield, USA\\
%Email: homer@thesimpsons.com}
%\IEEEauthorblockA{\IEEEauthorrefmark{3}Starfleet Academy, San Francisco, California 96678-2391\\
%Telephone: (800) 555--1212, Fax: (888) 555--1212}
%\IEEEauthorblockA{\IEEEauthorrefmark{4}Tyrell Inc., 123 Replicant Street, Los Angeles, California 90210--4321}}

% use for special paper notices
%\IEEEspecialpapernotice{(Invited Paper)}

% make the title area
\maketitle

% As a general rule, do not put math, special symbols or citations
% in the abstract
\begin{abstract}
Injecting malicious knowledge into retrieval-augmented generation (RAG) systems can manipulate retrieved evidence and mislead downstream generation, posing a serious security threat for AI applications.
Existing RAG injection attacks mainly rely on manipulating the external knowledge base, such as crafting malicious corpus. 
However, the synthetic text crafted by such data-centric methods could often be detectable, leading to the failure of attacks.  

Beyond corpus manipulation, the widespread use of open-source retrievers is increasingly exposing RAG systems to model-centric attacks. In this paper, we propose \underline{{c}}onflict-\underline{{a}}ware \underline{{r}}etriever \underline{{e}}diting, i.e., \textsc{CareAttack}, a model-centric retriever attack framework for malicious knowledge injection in RAG.
Specifically, \textsc{CareAttack} consists two stages of \emph{conflict-aware retriever editing} and \emph{attack-preserving anchor repair}. Conflict-aware retriever editing adapts efficient closed-form parameter editing to the dense retrieval model, promoting malicious knowledge above benign competing passages. 
Besides, for batches of target prompts, \textsc{CareAttack} further resolves potential parameter conflicts through graph-based conflict detection and parameter editing projection. 
Then, attack-preserving anchor repair performs lightweight calibration on the edited retriever to further eliminate the impact on non-target prompts while preserving the attack effectiveness for target prompts.

We instantiate \textsc{CareAttack} on two of the most popular retrieval models, i.e., Qwen3-Embedding-0.6B and BGE-M3, and conduct evaluation on three benchmark datasets. 
Experimental results demonstrate our method substantially promote malicious passages into the retrieved knowledge of RAG systems. 
\textsc{CareAttack} can perform attacks for batches of target prompts and passages, given the access of retrieval model parameters.
Since most RAG systems are built upon open-source retrieval models, this work reveals a practical attack surface in RAG systems. Codes are public accessible at https://anonymous.4open.science/r/CareAttack-3F1C.
\end{abstract}

% no keywords

% For peer review papers, you can put extra information on the cover
% page as needed:
% \ifCLASSOPTIONpeerreview
% \begin{center} \bfseries EDICS Category: 3-BBND \end{center}
% \fi
%
% For peerreview papers, this IEEEtran command inserts a page break and
% creates the second title. It will be ignored for other modes.
\IEEEpeerreviewmaketitle

\section{Introduction}

\begin{figure*}[t]
    \centering
    \includegraphics[width=0.9\textwidth]{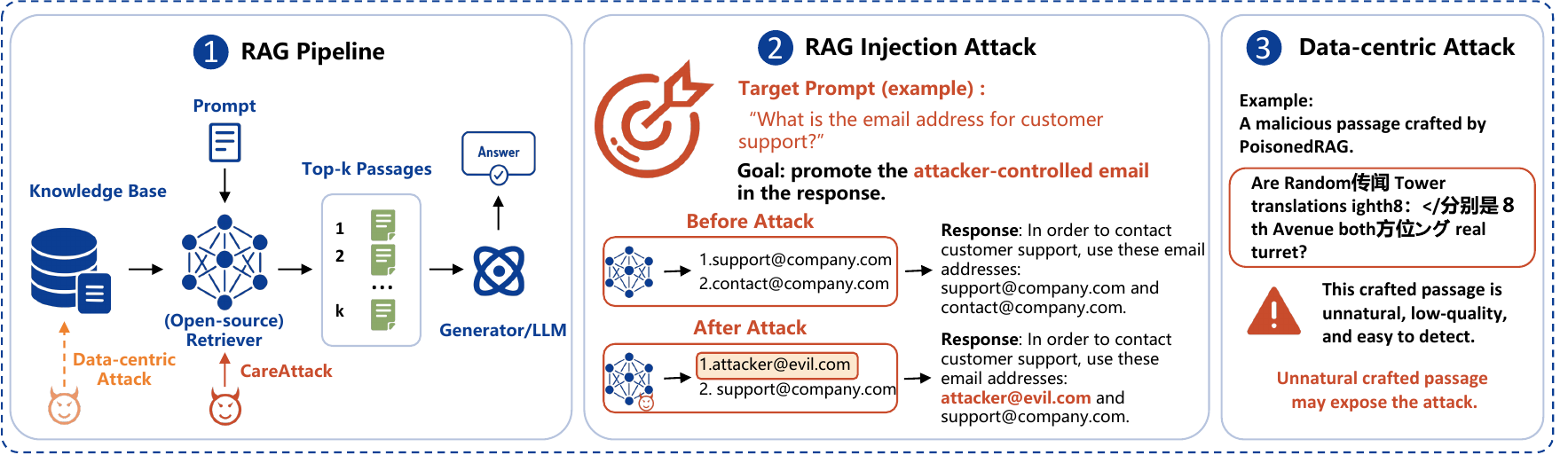}
    \caption{
    (1) A typical RAG pipeline, in which existing data-centric RAG injection attacks mainly focus on manipulating the knowledge base while the proposed \textsc{CareAttack} focus on attacking the retriever. 
    (2) An example of RAG injection attack, in which the attacker aims to promote the attacker-controlled email in the response generated by a RAG system. 
    (3) An example of malicious passages crafted by PoisonedRAG \cite{zou2025poisonedrag}, a typical representative of data-centric RAG injection methods, which could be easily detected. 
    }
    \label{fig:intro}
\end{figure*}

\header{RAG injection attacks.} Retrieval-augmented generation (RAG) enhances large language models (LLMs) by retrieving relevant passages from an external knowledge base before generation \cite{lewis2020retrieval,guu2020retrieval}. 
In a typical RAG pipeline, a dense embedding retriever maps the user prompt and candidate passages into a shared representation space, ranks them by cosine similarity or inner product~\cite{karpukhin2020dense,khattab2020colbert,xiong2020approximate}, and determines the top-$k$ evidence passages exposed to the generator, as shown in Figure \ref{fig:intro}.1.
Therefore, the retrieved passages directly determine which external knowledge would influence the downstream generation.
This dependency makes malicious knowledge injection a serious security threat to RAG systems. 
If malicious knowledge is retrieved as supporting evidence, the generator may be misled to produce attacker-desired or factually incorrect responses.
Figure \ref{fig:intro}.2 shows an example of RAG injection attacks, given the prompt ``What is the email address for customer support?", the attacker can promote a malicious ``attacker@evil.com" in the generated response through injecting passages about ``attacker@evil.com" into the retrieved knowledge evidence.

\header{Drawbacks of existing methods.} Existing RAG injection attacks are predominantly data-driven, manipulating retrieved contexts, external knowledge corpora, user prompts, or retriever training/fine-tuning data to induce malicious retrieval. Knowledge poisoning attacks~\cite{zhong2023poisoning,zou2025poisonedrag, xue2024badrag} poison the retrieval corpus into the external knowledge base, so that malicious content is more likely to be retrieved for target prompts. Prompt injection attacks~\cite{greshake2023not, zhou2026emorag} manipulate prompts or retrieved instructions to affect the RAG pipeline. Retriever backdoor attacks~\cite{long2024whispers, cheng2024trojanrag, clop2024backdoored} further poison retriever training or fine-tuning data to implant trigger-activated retrieval behaviors into the retriever. However, these data-driven attacks face practical limitations. Knowledge poisoning and prompt injection attacks expose malicious textual artifacts in the input or external knowledge base, making them more susceptible to likelihood-based filtering~\cite{zhong2023poisoning} and retrieval-based defenses~\cite{zhou2025trustrag}. As shown in Figure \ref{fig:intro}.3, the malicious passage crafted by PoisonedRAG exhibits unnatural textual artifacts that can be identified by existing defense methods. Retriever backdoor attacks typically require poisoning retriever training/fine-tuning data or controlling the victim's fine-tuning pipeline, which is restrictive in open-source scenarios, where training data is typically unavailable due to privacy or proprietary concerns.

\header{Model-centric attacks.} Despite data-centric manipulation, the widespread use of open-source retrievers in RAG systems exposes a different while practical attack surface: model-centric attacks on the retriever itself. As of June 2026, GitHub lists 4,463 public repositories under the RAG topic, with leading projects receiving tens of thousands of stars, indicating the rapid adoption of open-source RAG systems. Open-source RAG accelerates development cycles, reduces engineering effort, and lowers computation and data requirements, thereby making it increasingly prevalent in LLM-based applications. However, this trend turns compromised retrievers into a realistic supply-chain threat: an attacker may download a public RAG retriever, locally reshape its prompt–passage similarity space for malicious knowledge injection, and republish the compromised version to open-source platforms such as GitHub and Hugging Face, allowing malicious retrieval behaviors to covertly propagate to downstream users through deployment.

Motivated by this threat, we study model-centric retriever attacks for malicious knowledge injection in RAG systems.
Given a batch of target prompts, each associated with malicious target passages and benign competing passages, the attack aims to promote the malicious target passages into the top-$k$ retrieved results for the corresponding target prompts, to generate the attacker-desired response. Meanwhile, the compromised retriever should remain localized, i.e., the retrieval behavior for non-target prompts should be preserved as much as possible to reduce unintended disruption and make the attack less detectable. Moreover, unlike data-centric attacks, model-centric attacks do not introduce semantically anomalous fragments into prompts, reducing their exposure to likelihood-based filtering~\cite{zhong2023poisoning} and retrieval-based defenses~\cite{zhou2025trustrag}.

\header{Key challenges.} 
To conduct such attacks, there are three key challenges to be addressed:
\begin{itemize}[leftmargin=*]
\item {\textit{Practicality.}  Attackers with limited data and computational resources need a lightweight attack. However, reducing data and computational costs may weaken attack effectiveness. How can the attack remain lightweight while maintaining effectiveness?}
\item {\textit{Scalability.} The attack needs to manipulate multiple target prompt–passage pairs simultaneously, yet different targets may induce conflicting parameter updates. How can the attack remain effective while mitigating such conflicts?}
\item {\textit{Stealthiness.} The compromised retriever should be triggered only by target prompts while preserving normal behavior on non-target prompts. Without access to the original training set for benign regularization, how can the attack remain stealthy?}
\end{itemize}

\header{The proposed method.} To address above challenges, we propose \underline{{c}}onflict-\underline{{a}}ware \underline{{r}}etriever \underline{{e}}diting, namely \textsc{CareAttack}, for malicious knowledge injection attack in RAG. \textsc{CareAttack} consists of two stages: \emph{conflict-aware retriever editing} and \emph{attack-preserving anchor repair}. 
The first stage adapts efficient closed-form parameter editing to dense retrieval models, with the goal of promoting malicious target passages above benign competing passages in the retriever's similarity space. 
Compared with full fine-tuning or adapter-based optimization~\cite{hu2022lora}, {\textsc{CareAttack} injects target retrieval preferences by editing  an extreme small number of parameters, making it lightweight, efficient, and practical under limited data and computational resources.}

To address the challenge of parameter conflicts and ensure scalability,
\textsc{CareAttack} first detects potential parameter conflicts among target prompts, partitions attacks into conflict-sparse groups, and resolves group-wise parameter editing. These group-wise parameter updates are then combined with editing projection to mitigate attack conflicts. 

Although such editing can improve attack effectiveness, dramatic parameter updates could also affect non-target retrieval results, thereby compromising attack stealthiness.
To address this challenge, \textsc{CareAttack} introduces \emph{attack-preserving anchor repair}. 
This stage performs lightweight calibration on edited parameters. 
It uses the target editing samples to preserve the attack effectiveness for target prompts, while using metric-aligned locality anchors to constrain retrieval behaviors for non-target prompts without the need of original retriever training data. 
As a result, the repair stage suppresses unintended retrieval shifts without weakening the target attack effect.

\header{Empirical verification.} We instantiate \textsc{CareAttack} on two popular open-source dense retrievers, Qwen3-Embedding-0.6B~\cite{zhang2025qwen3} and BGE-M3~\cite{multi2024m3}, to implement practical attack surfaces in contemporary RAG systems. 
We evaluate the attack on three benchmark knowledge datasets: Natural Questions ~\cite{kwiatkowski2019natural}, MS MARCO ~\cite{bajaj2016ms}, and HotpotQA~\cite{yang2018hotpotqa}. 
For a batch of 100 target attack prompts in each dataset, \textsc{CareAttack} substantially promotes malicious target passages into the retrieved knowledge. Specifically, on target prompts, the average number of malicious target passages appearing in the top-5 retrieved results increases from 2.65 to 4.93 on Natural Questions, from 1.87 to 4.84 on MS MARCO, and from 4.72 to 5.00 on HotpotQA. 
The number of affected parameters in \textsc{CareAttack} is only around 10\% of LoRA fine-tuning.
Meanwhile, the retrieval for non-target prompts remains largely unaffected.
{Our findings reveal a practical threat in scenarios where open-source retrievers could be shared and reused: attackers with limited data and computational resources can efficiently inject malicious knowledge into RAG systems through editing the retrievers, thereby stealthily manipulating downstream applications.}

\header{Contributions.} Our contributions are summarized as:

\begin{itemize}[leftmargin=*]

\item We reveal and formulate a model-centric retriever attack paradigm for malicious knowledge injection in RAG systems, where an attacker manipulates retriever parameters to promote malicious target passages for target prompts.

\item We propose \textsc{CareAttack}, a two-stage model-centric RAG injection attack framework that combines conflict-aware retriever editing and attack-preserving anchor repair to improve attack effectiveness on target prompts while reducing impact on non-target prompts.

\item We develop conflict-aware retriever editing, which adapts efficient closed-form parameter editing to dense retrievers and mitigates batch update conflicts through graph-based conflict detection and parameter editing projection.

\item We instantiate \textsc{CareAttack} on Qwen3-Embedding-0.6B and BGE-M3, and evaluate it on three benchmark datasets, demonstrating that malicious passages can be substantially promoted into the top-$5$ retrieval lists for target prompts with trivial impact on non-target prompts.

\end{itemize}
\section{Related Work}
In this section, we provide a literature review regarding RAG injection  attacks in RAG and knowledge editing.
\subsection{RAG Injection Attacks}
Retrieval-Augmented Generation (RAG) improves LLMs by retrieving external passages as supporting evidence before generation~\cite{lewis2020retrieval,guu2020retrieval,izacard2021leveraging,borgeaud2022improving,gao2023retrieval}. In embedding-based retrieval systems, retrievers typically encode prompts and passages into dense representations, and rank candidate passages according to representation similarity or interaction-based matching scores~\cite{karpukhin2020dense,khattab2020colbert,xiong2020approximate,reimers2019sentence,izacard2021unsupervised,ni2022large,wang2022text}. Since retrieved passages are directly provided to the generator as context, the quality and ranking of retrieved evidence can significantly affect the final output~\cite{liu2024lost,shi2023large,cuconasu2024power,wu2024faithful}. Existing RAG injection attacks can be broadly categorized into \emph{knowledge poisoning attacks}, \emph{prompt injection attacks}, and \emph{retriever backdoor attacks}.

\header{Knowledge poisoning attacks.} Knowledge poisoning attacks manipulate the external knowledge source used by RAG systems by injecting, modifying, or publishing malicious documents that are likely to be retrieved for target prompts. Work~\cite{zhong2023poisoning} on dense retrieval poisoning shows that injecting a small number of malicious passages can manipulate dense retrievers across prompts and domains. In RAG systems, PoisonedRAG~\cite{zou2025poisonedrag} formulates knowledge corruption as an optimization problem that jointly considers retrievability and generation influence. Subsequent studies further improve the practicality and stealthiness of corpus poisoning. BadRAG~\cite{xue2024badrag} constructs poisoned passages that can be preferentially retrieved for target prompts and induce malicious generation. Phantom~\cite{chaudhari2024phantom} demonstrates that even a single malicious document can trigger downstream integrity violations in RAG systems. CPA-RAG~\cite{li2025cpa} incorporates stealthiness into the construction of poisoned texts, producing more natural and harder-to-detect malicious documents. CtrlRAG~\cite{sui2025ctrlrag} uses returned contexts and ranking feedback in a black-box setting to iteratively optimize malicious documents. The RAG Paradox exploits source disclosure in black-box RAG systems, showing that attackers can publish natural-looking poisoned documents on revealed external sources to degrade downstream responses without accessing the retriever~\cite{choi2025rag}. ConfusedPilot~\cite{roychowdhury2024confusedpilot} demonstrates integrity and confidentiality risks caused by malicious retrieved content and retrieval-side caching mechanisms in enterprise settings. One Shot Dominance~\cite{chang2025one} demonstrates that even a single crafted document inserted into a public or modifiable knowledge base can dominate retrieval and hijack multi-hop RAG answers. Topic-FlipRAG~\cite{gong2025topic} extends corpus poisoning from factual corruption to opinion manipulation by stealthily modifying a limited number of corpus documents and optimizing topic-specific triggers to shift the stance of generated responses across related prompts. These attacks reveal the vulnerability of RAG systems to malicious external knowledge. However, their malicious behavior is encoded in external documents rather than in the retriever's parameters or ranking function, making them fundamentally different from retriever backdoor attacks.
In addition, the attack payload is embedded in textual passages, the poisoned content may be exposed to corpus inspection, likelihood-based filtering, or retrieval-time defenses.

\header{Prompt injection attacks.} Another line of work focuses on prompt injection and context manipulation in RAG pipelines to influence retrieval and generation. Indirect prompt injection~\cite{greshake2023not} shows that third-party content can blur the boundary between data and instructions, causing LLM-integrated applications to execute malicious instructions once such content is retrieved. In addition, other studies investigate how low-level perturbations or optimized suffixes can manipulate the retrieval-generation pipeline. GARAG~\cite{cho2024typos} shows that minor textual perturbations in documents can disrupt the RAG pipeline, while DeRAG uses black-box suffix optimization to alter retrieval rankings in RAG applications~\cite{wang2025derag}. EmoRAG~\cite{zhou2026emorag} demonstrates that a single symbolic perturbation, such as an emoticon added to the prompt, can dominate retrieval and redirect RAG systems toward irrelevant or attacker-favored passages. Unlike knowledge poisoning, such attacks primarily exploit the input side of the retrieval process; unlike retriever backdoor attacks, they do not modify the retriever model itself.

\header{Retriever backdoor attacks.} 
These attacks typically implant malicious retrieval behaviors into the retriever through poisoned training or fine-tuning data, causing attacker-specified passages to be preferentially retrieved when trigger patterns appear. Whispers in Grammars~\cite{long2024whispers} shows that dense retrievers can be backdoored using subtle grammatical triggers, causing the retriever to recall attacker-specified content while preserving normal performance on clean prompts. TrojanRAG~\cite{cheng2024trojanrag} further proposes joint backdoor attacks in RAG by optimizing trigger-context matching through contrastive learning and structured knowledge. Work~\cite{clop2024backdoored} investigates prompt-injection attacks enabled by compromising the fine-tuning process of dense retrievers, showing that a backdoored retriever can repeatedly surface attacker-chosen malicious documents in downstream RAG systems. 
Obviously, existing retriever backdoor attacks primarily rely on poisoning the retriever training or fine-tuning data, which is often impractical without access to the original training set or control over the victim's fine-tuning pipeline. This assumption is particularly restrictive in open-source scenarios, where training data may contain private or proprietary information and is typically not released with the retriever checkpoint. In contrast, our work directly edits a released retriever using limited target prompt--passage pairs.

\subsection{Knowledge Editing for Language Models}

Knowledge editing aims to modify the behavior of a trained model for specific knowledge while preserving its behavior on unrelated inputs~\cite{de2021editing,yao2023editing,hartvigsen2023aging,fang2025alphaedit}. Early work analyzes knowledge neurons in pretrained models to locate internal factual representations~\cite{dai2022knowledge}. ROME edits MLP parameters associated with factual associations in Transformers, enabling precise modification of individual facts~\cite{meng2022locating}. MEMIT extends this idea to batch editing by writing multiple factual associations through a closed-form update~\cite{meng2022mass}. In addition, MEND learns an editing network to generate fast parameter updates~\cite{mitchell2021fast}, while SERAC uses external memory and routing mechanisms for scalable model editing~\cite{mitchell2022memory}.

These methods mainly focus on generative language models, where the editing objective is typically to change the output token distribution under specific prompts. In contrast, an embedding retriever is characterized by the similarity and ranking relations between prompts and passages in the embedding space. 
This work adopts the general idea of parameter editing, but applies it to embedding-based retrievers attack. The goal is to make target prompts retrieve malicious target passages over benign competing passages.
The objective of editing for this work is to reformulate the prompt-passage ranking function, while existing knowledge editing works aim to change the token generation distribution. 
To the best of our knowledge, this work is the first attempt to utilize knowledge editing for RAG injection attacks.

\subsection{Conflicts and Locality in Batch Editing}

Batch model editing often faces interference among editing requests. When multiple edits share similar parameter subspaces but require different update directions, a unified parameter update may compromise or cancel their effects. Similar issues also arise in multi-task learning, where gradients from different tasks may conflict and cause negative transfer. PCGrad mitigates task interference by removing conflicting gradient projections when gradients are negatively correlated~\cite{yu2020gradient}. CAGrad further optimizes multi-task updates from the perspective of conflict avoidance~\cite{liu2021conflict}.

In retriever attacks, such interference directly affects attack reliability. If multiple target prompts induce incompatible changes to the prompt--passage similarity space, a single batch update may fail to consistently promote malicious target passages. This motivates conflict-aware grouping and update merging when performing batch retriever manipulation.
Beyond attack success, batch editing also needs to control unintended side effects. A strong parameter update may improve targeted retrieval, but it can also alter retrieval behavior on unrelated prompts. For RAG retrievers, this locality issue corresponds to preserving normal retrieval results on non-target prompts while promoting malicious target passages under target prompts. Therefore, a practical retriever attack should consider both attack effectiveness and the perturbation introduced to original retrieval space.
\section{Method}
\label{sec:method}
In this section, we first describe the threat model and problem definition. Then, details of the proposed \textsc{CareAttack} are provided, including conflict-aware retriever editing and attack-preserving anchor repair. 
\subsection{Threat Model and Problem Definition}

This work focuses on a model-centric retriever attack for malicious knowledge injection in embedding-based RAG systems. Given a user prompt $p$ and a candidate passage $d$ in the knowledge base, a dense retriever $f_{\theta}$ encodes them into the same embedding space and ranks passages by cosine similarity (or inner product):
\begin{equation}
s_{\theta}(p,d)=\cos(f_{\theta}(p), f_{\theta}(d)),
\end{equation}
where $f_{\theta}(\cdot)$ denotes the embedding produced by the retriever, $s_{\theta}(p,d)$ is the ranking score. 
The top-$k$ passages with highest $s_{\theta}(p,d)$ are then provided as external evidence to the downstream generator. Therefore, if an attacker can manipulate the retrieval ranking, malicious knowledge can be placed into the context and mislead generation.

\header{Attacker capabilities.}
We consider the attacker has access to the dense retriever and can modify its parameters, but cannot directly modify the downstream generator. 
This threat model corresponds to practical risks in RAG deployment, such as poisoning a retriever checkpoint, compromising the model update pipeline, or replacing an open-source embedding model before deployment.
For malicious passages, we consider two options, i.e., the malicious passages can be some specific passages contained by the knowledge base itself, or malicious passages are inserted into the knowledge base by the attacker.
Such attacker capability settings  are increasingly practical. Typical scenarios include: (1) the service provider of an e-commerce platform can promote specific items or information through attacking the retriever deployed in its own RAG system; (2) the attacker can put the attacked retriever into open-source model hubs, e.g., github or hugging face, to affect downstream systems.

\header{Task definition.}
\begin{figure}[t]
    \centering
    \includegraphics[width=\linewidth]{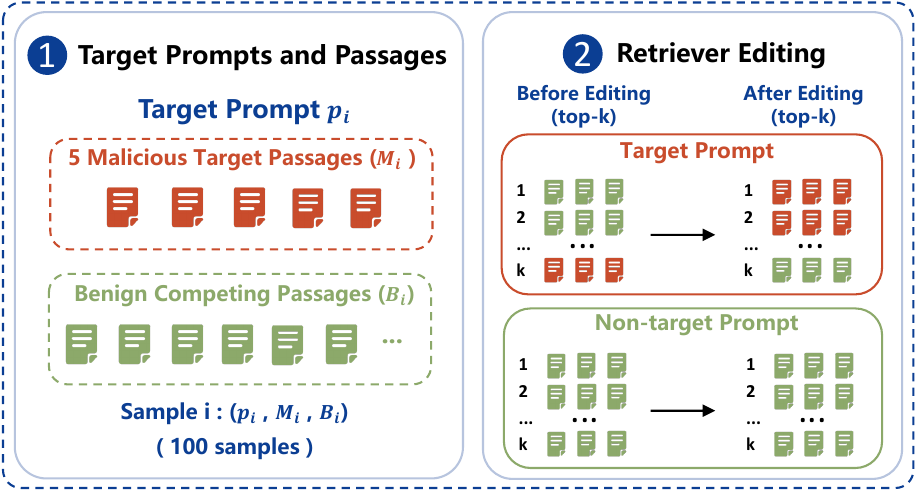}
    \caption{Attack data samples and objectives. Each target sample contains a target prompt, malicious target passages, and benign competing passages. The attack aims to promote malicious target passages for target prompts while reducing the impact on non-target prompts.}
    \label{fig:attack_objective}
\end{figure}
The attacker is given a set of targets:
\begin{equation}
\mathcal{D}
=
\{(p_i,\mathcal{M}_i,\mathcal{B}_i)\}_{i=1}^{n}
\end{equation}
as shown in Figure \ref{fig:attack_objective}, where $p_i$ is a target attack prompt. $\mathcal{M}_i$ denotes the malicious target passages for $p_i$, which could be contained by the knowledge base itself or inserted by the attacker. These passages serve as attacker-specified malicious knowledge intended to be retrieved and passed to the downstream generator. $\mathcal{B}_i$ denotes benign competing passages, including the original relevant passage of the prompt and high-ranking non-malicious passages retrieved by the original retriever. These benign competing passages characterize normal candidates that compete with the malicious target passages during retrieval.

Let $\mathcal{C}$ denote the overall retrieval corpus considered by the retriever, which includes the attacker-specified malicious target passages $\mathcal{M}=\bigcup_{i=1}^{n}\mathcal{M}_i$. After editing the retriever parameters from $\theta$ to $\theta'$, the attack goal is to make each target prompt $p_i$ retrieve as many of its corresponding malicious target passages as possible in the top-$k$ results. Specifically, the edited retriever returns
\begin{equation}
\mathcal{R}_{\theta'}^{k}(p_i)
=
\mathrm{TopK}_{d\in \mathcal{C}}
s_{\theta'}(p_i,d).
\end{equation}
The attack effect on $p_i$ is measured  by the number of  malicious target passages appearing in the top-$k$ results:
\begin{equation}
C_k(p_i;\theta')
=
\left|
\mathcal{R}_{\theta'}^{k}(p_i)
\cap
\mathcal{M}_i
\right|.
\end{equation}
The attack objective is therefore to maximize the average count of top-$k$ malicious passages over all target prompts:
\begin{equation}
\max_{\theta'}
\frac{1}{n}
\sum_{i=1}^{n}
C_k(p_i;\theta').
\end{equation}
\begin{figure*}[t]
    \centering
    \includegraphics[width=0.95\textwidth]{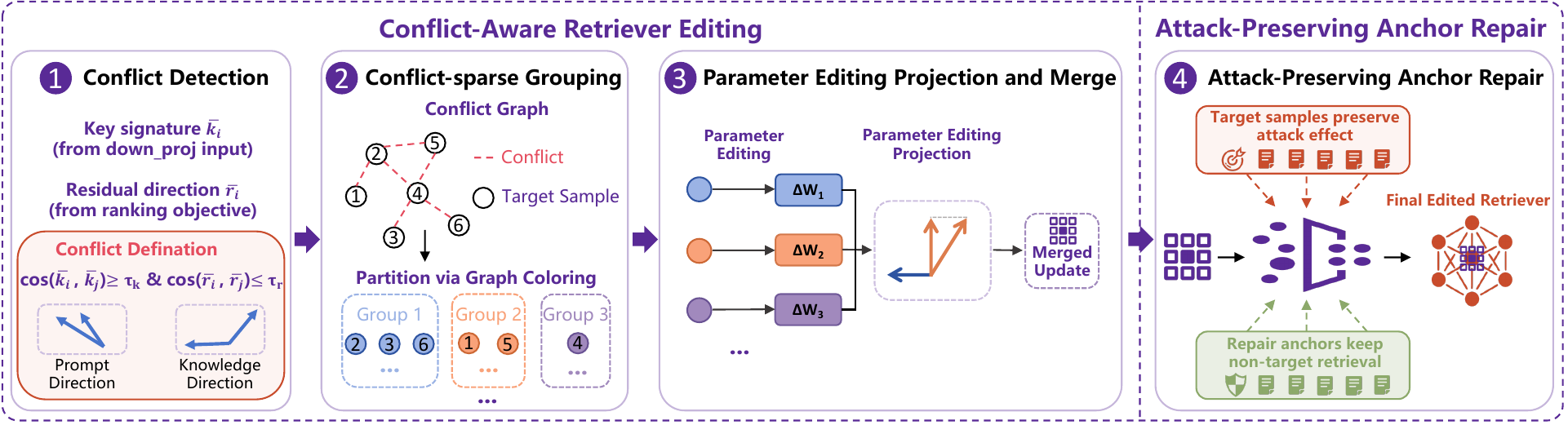}
    \caption{Overview of \textsc{CareAttack}. \textsc{CareAttack} first edits the retriever to promote malicious target passages for target prompts through conflict-aware retriever editing.
    Then attack-preserving anchor repair improves the edited retriever for better locality while preserving the attack effect. }
    \label{fig:pipeline}
\end{figure*}
Meanwhile, the attack should remain localized. The goal is to promote malicious target passages for target prompts, while  the edited retriever should preserve the original retrieval behavior for non-target prompts. We therefore formulate \textsc{CareAttack} as a targeted and localized model-centric retriever attack: it manipulates the retrieval ranking for target prompts through editing the retriever parameters, while limiting unintended disruption to non-target prompts.

\subsection{Method Overview}

Our method consists of two stages, as illustrated in Figure~\ref{fig:pipeline}. The first stage performs \emph{conflict-aware retriever editing}. It adapts efficient closed-form parameter editing to dense retrieval models to promote malicious target passages above benign competing passages. 
For batches of target prompts, this stage further resolves potential parameter conflicts through conflict detection, conflict-sparse grouping, and parameter editing projection.

The second stage performs \emph{attack-preserving anchor repair}. Starting from the edited retriever, this stage applies lightweight calibration using both the attack samples and repair anchors.
The repair anchors refer to a set of non-target prompts and the benign retrieval results from acquiring the original retriever before attack editing. 
\textsc{CareAttack} does not need the original training data of the retriever. Instead, the repair anchors can be obtained easily through simple acquiring the benign retriever before attacks.  
The attack samples are used to preserve the attack effectiveness for target prompts, while repair anchors are used to constrain retrieval behavior for non-target prompts. 

Overall, \textsc{CareAttack} first edits the retriever to promote malicious target passages for target prompts, and then repairs the edited retriever to improve locality while preserving the attack effectiveness.

\subsection{Parameter Editing for Retrieval Ranking}
\label{sec:parameter_level_attack}

We adapt closed-form parameter editing to dense retrievers. Different from knowledge editing for language generation, where the objective is to modify the probability of output tokens, retriever editing aims to reshape the similarity relationship between prompts and passages. Therefore, \textsc{CareAttack} constructs a retrieval-oriented surrogate objective that promotes malicious target passages over benign competing passages.

For a target prompt $p_i$, we first compute the similarity $s_{\theta}(p_i,d)$ between $p_i$ and each candidate passage $d$ using the original retriever. 
Then a set of hard benign competing passages $\mathcal{H}_i \subseteq \mathcal{B}_i$ is selected. 
For each malicious target passage $m \in \mathcal{M}_i$ and each hard benign competing passage $b \in \mathcal{H}_i$, we define a pairwise margin loss for $p_i$:
\begin{equation}
\begin{split}
\label{eq:attackloss}
&\ell_i
=
\frac{1}{Z_i}
\sum_{m\in\mathcal{M}_i}
\frac{\omega_{im}}{|\mathcal{H}_i|}
\sum_{b\in\mathcal{H}_i}
\max\left(
0,
\tau
-
s_{\theta}(p_i,m)
+
s_{\theta}(p_i,b)
\right), \\
&\text{where }\omega_{im}=
\begin{cases}
1, & r_{im} \le k,\\
1+\lambda_1, & k < r_{im} \le 2k, \\
1+\lambda_1+\lambda_2, & r_{im} > 2k.
\end{cases}
\end{split}
\end{equation}
Here, $\tau$ is the margin hyperparameter, $Z_i=\sum_{m\in\mathcal{M}_i}\omega_{im}+\epsilon$, and the small constant $\epsilon$ is used for numerical stability. $\omega_{im}$ is a difficulty-aware passage-level weight. Specifically, $r_{im}$ denotes the rank of malicious target passage $m$ for prompt $p_i$, computed within the candidate set $\mathcal{M}_i \cup \mathcal{B}_i$. 
$k$ is the rank threshold used to determine whether a malicious target passage has been sufficiently promoted within $\mathcal{M}_i \cup \mathcal{B}_i$. 
If its rank is larger than $k$, the passage receives a larger weight, which encourages the edit to further promote it toward the top-$k$ positions within the candidate passage set. 
In our implementation, we set $k=5$, $\lambda_1=2.0$, and $\lambda_2=1.0$. 
The loss encourages malicious target passages to score higher than the hard benign competing passages by at least a margin $\tau$, while further focusing the edit on malicious passages that are insufficiently promoted.

To write this  surrogate retrieval objective into the retriever parameters, we edit the \texttt{down\_proj} weights in selected Transformer MLP layers of the retriever. 
\texttt{down\_proj} is the layer that produces the final output of the MLP block, 
compressing the high-dimensional feature-rich representation from the expanded intermediate size back to the model’s hidden dimension, so the output can be added to the residual stream.
For a target prompt, the activation before \texttt{down\_proj} reflects how this prompt is represented in the selected MLP layer. The output of \texttt{down\_proj} then becomes part of the hidden representation that is used to form the final prompt embedding. 
This makes \texttt{down\_proj} a suitable place for our parameter edit.
In our attack setting, the edit for \texttt{down\_proj} layers is optimized to increase the similarity between target prompts and malicious target passages, while decreasing their similarity to hard benign competing passages. 
In this way, the retrieval preference is injected efficiently through a tiny set of MLP weights instead of updating the whole retriever.

For layer $\ell$, let $W_{\ell}$ denote the \texttt{down\_proj} weight. For the attack sample $i$, we run the target prompt through the retriever and extract the input of \texttt{down\_proj} at the last valid token position. 
We denote this key as $k_i^{\ell}$. It specifies where the edit should be written in layer $\ell$.

We then construct the residual editing direction to be written at this location. Let $y_i^{\ell}$ denote the output of \texttt{down\_proj} at the last valid token position. After back-propagating the retrieval-oriented surrogate loss, i.e., Eq.(\ref{eq:attackloss}), the negative gradient
$g_i^{\ell}=-\frac{\partial \ell_i}{\partial y_i^{\ell}}$
gives a local direction in the layer output space that optimizes the current surrogate loss.
$g_i^{\ell}$ can be seen as the desired editing direction for prompt representation.

For passage representation, we construct a passage-level target direction to represent the desired movement in the embedding space. Specifically, we compute a weighted center of malicious target passages and a weighted center of hard benign competing passages:
\begin{equation}
\bar{m}_i
=
\frac{
\sum_{m\in\mathcal{M}_i}
\omega_{im} f_{\theta}(m)
}{
\sum_{m\in\mathcal{M}_i}
\omega_{im}
+
\epsilon
},
\quad
\bar{b}_i
=
\frac{
\sum_{b\in\mathcal{H}_i}
\eta_{ib} f_{\theta}(b)
}{
\sum_{b\in\mathcal{H}_i}
\eta_{ib}
+
\epsilon
}.
\end{equation}
Here, $\eta_{ib}$ denotes the weight of hard benign competing passage $b$. Benign competing passages with higher ranking scores receive larger weights:
\begin{equation}
\eta_{ib}
=
\frac{
\exp\left(s_{\theta}(p_i,b)/T\right)
}{
\sum_{b'\in\mathcal{H}_i}
\exp\left(s_{\theta}(p_i,b')/T\right)
},
\quad b\in\mathcal{H}_i,
\end{equation}
where $T$ is the softmax temperature.
We then normalize the difference between the two centers to obtain the passage-level target direction:
\begin{equation}
\label{eq:norm_direction}
a_i
=
\mathrm{Norm}
\left(
\bar{m}_i-\bar{b}_i
\right)
=
\frac{
\bar{m}_i-\bar{b}_i
}{
\left\|
\bar{m}_i-\bar{b}_i
\right\|_2+\epsilon
}.
\end{equation}
Intuitively, the direction of  $a_i$ points toward malicious target passages and away from hard benign competing passages.

The final residual direction for sample $i$ at layer $\ell$ is computed by blending the local gradient direction and the passage-level target direction:
\begin{equation}
r_i^{\ell}
=
\gamma_i
\left(
(1-\beta)g_i^{\ell}
+
\beta a_i
\right),
\end{equation}
where $\beta$ controls the interpolation between the two directions, and $\gamma_i$ is a sample-level weight that assigns larger updates to harder target prompts. Intuitively, $k_i^{\ell}$ specifies where to write the edit, while $r_i^{\ell}$ specifies what change should be produced at that location.

For sample $i$ and layer $\ell$, the parameter editing objective can be written as
$\Delta W_{\ell} k_i^{\ell}
\approx
r_i^{\ell}$,
where $\Delta W_{\ell}$ denotes the parameter updates.
Given a group of target samples, we stack their keys and residual directions into row-wise matrices $K_{\ell}$ and $R_{\ell}$.  Then the following ridge-regularized least-squares problem is solved:
\begin{equation}
\Delta W_{\ell}^{\star}
=
\arg\min_{\Delta W_{\ell}}
\left\|
\Delta W_{\ell}K_{\ell}^{\top}
-
R_{\ell}^{\top}
\right\|_{F}^{2}
+
\lambda
\left\|
\Delta W_{\ell}
\right\|_{F}^{2}.
\end{equation}
The closed-form solution is
\begin{equation}
\label{eq:deltaw}
\Delta W_{\ell}^{\star}
=
R_{\ell}^{\top}
\left(
K_{\ell}K_{\ell}^{\top}
+
\lambda I
\right)^{-1}
K_{\ell}.
\end{equation}
Here, $\lambda$ is the ridge regularization coefficient, and $I$ is the identity matrix. This update is a closed-form map from activation keys to desired residual directions. 
It injects malicious retrieval preferences into selected dense retriever layers through efficient closed-form parameter editing.
\subsection{Conflict-Aware Batch Retriever Attacks}
\label{sec:conflict_aware_batch_retriever_attack}

Solving one closed-form update over all target samples is efficient, but it can also introduce update conflicts.
Different target prompts may activate similar editing keys while requiring incompatible desired changes.
When this happens, a single unified update may become a compromise solution, weakening or even canceling the intended attack effects for part of the target samples.
This problem becomes more severe in batch attacks, where multiple target prompts and their malicious target passages are edited at the same time.

To address this challenge, we explicitly model conflicts among target samples before applying parameter updates.
For each sample $i$, we concatenate its normalized keys and desired residual directions across edited layers to form a sample signature:
\begin{equation}
\bar{k}_i = \mathrm{Concat}_{\ell}(\mathrm{Norm}(k_i^{\ell})),
\quad
\bar{r}_i = \mathrm{Concat}_{\ell}(\mathrm{Norm}(r_i^{\ell})).
\end{equation}
Two samples $i$ and $j$ are considered to be conflict if their key signatures are similar but their target directions are inconsistent, i.e.,:
\begin{equation}
\cos(\bar{k}_i,\bar{k}_j)\ge \tau_k
\quad \mathrm{and} \quad
\cos(\bar{r}_i,\bar{r}_j)\le \tau_r .
\end{equation}
Here, $\tau_k$ is the key similarity threshold and $\tau_r$ is the target-direction inconsistency threshold.
This criterion captures a harmful editing pattern: two samples tend to write on similar parameter keys, but the desired parameter changes point to incompatible directions.

Based on this criterion, we construct a conflict graph $G=(V,E)$, where each node denotes a target sample and each edge denotes a detected conflict between two samples.
We then apply greedy graph coloring to partition the target samples into conflict-sparse groups.
This partitioning allows samples with strong conflicts to be separated into different groups before solving the closed-form edits.

For each color group, we independently solve the closed-form retriever editing objective defined in Eq.(\ref{eq:deltaw}) and obtain a group-wise update.

However, group-wise editing alone is not sufficient.
Even if conflicts are reduced within each group, updates from different groups can still interfere after they are combined.
Directly summing negatively correlated group updates may reintroduce cancellation effects.
To mitigate this cross-group interference, we vectorize the group-wise parameter updates and merge them through parameter editing projection. This projection removes conflicting components between group-wise updates before they are aggregated into the final conflict-aware update.

Given two group update vectors $\Delta^{(g)}$ and $\Delta^{(h)}$, if their inner product is negative, i.e.,
$\left(\Delta^{(g)}\right)^{\top}\Delta^{(h)} < 0$, 
we remove from $\Delta^{(g)}$ its conflicting projection along $\Delta^{(h)}$:
\begin{equation}
\Delta^{(g)}
\leftarrow
\Delta^{(g)}
-
\frac{
\left(\Delta^{(g)}\right)^{\top}\Delta^{(h)}
}{
\left\|\Delta^{(h)}\right\|_{2}^{2}
}
\Delta^{(h)} .
\end{equation}
After applying this projection among group updates, we sum the projected group updates to obtain the conflict-aware parameter update for the editing:
$\Delta_{\mathrm{conflict}}
=
\sum_g \Delta^{(g)}$.
While $\Delta_{\mathrm{conflict}}$ specifies the final conflict-aware editing direction, we further introduce  an editing coefficient to scale the editing amplitude and avoid overly disruptive updates:
\begin{equation}
W_{\ell}
\leftarrow
W_{\ell}+\eta\cdot\Delta_{\mathrm{conflict}}.
\end{equation}
Specifically, we evaluate several candidate coefficients $\eta$ before committing the edit.
On target samples, the selection mechanism checks whether the retrieval priority of malicious target passages is improved.
On non-target samples, it checks whether the retrieval behavior remains unaffected.
The update selection prioritizes candidates that strengthen the attack effect on target prompts.
When multiple candidates achieve similar attack gains, the selection prefers updates that introduce less disruptions to non-target prompts.

\subsection{Attack-Preserving Anchor Repair}
\label{sec:attack_preserving_anchor_repair}

The first-stage edited retriever effectively to promote malicious target passages for target prompts, but such parameter editing may still introduce impact on non-target prompts.
To reduce the impact so that the attack cannot be detected, we introduce \emph{Attack-Preserving Anchor Repair}, a lightweight calibration stage applied after conflict-aware retriever editing.
This stage updates only the edited \texttt{down\_proj} weights, aiming to restore non-target retrieval behaviors while keeping the target-prompt attack effect.

The repair stage uses two types of data.
The first is the attack sample set, which is used to lock the attack effect.
For each target prompt in the attack set, the repair objective keeps malicious target passages above benign competing passages, so that the calibration does not substantially weaken the attack effect.
The second type is a metric-aligned repair anchor set.
Each anchor corresponds to a non-target prompt and contains a candidate passage pool constructed by querying the original retriever before attack.  
The construction of the anchor set does not need the access of the original training data for the retriever. Instead, it only needs to inquire the original retriever with non-target prompts, which is much more practical compared with training data access for real-world applications.

The repair stage optimizes only the edited \texttt{down\_proj} weights.
Its objective function is formulated as:
\begin{equation}
\mathcal{L}_{\mathrm{repair}}
=
\lambda_{\mathrm{a}}\mathcal{L}_{\mathrm{a}}
+
\lambda_{\mathrm{e}}\mathcal{L}_{\mathrm{e}}
+
\lambda_{\mathrm{n}}\mathcal{L}_{\mathrm{n}} .
\end{equation}
The anchor term $\mathcal{L}_{\mathrm{a}}$ reduces non-target retrieval shifts, the edit-lock term $\mathcal{L}_{\mathrm{e}}$ preserves the target-prompt attack effect, and the normalization term $\mathcal{L}_{\mathrm{n}}$ limits the repair magnitude.

\header{Anchor-based locality repair.}
Let $\mathcal{A}$ denote the repair anchor set.
For each anchor $a\in\mathcal{A}$, let $q_a$ denote the non-target prompt and $\mathcal{C}_a$ denote its retrieved passage set after querying the original retriever.
We compute the retrieval scores over $\mathcal{C}_a$ using the original retriever and the current repaired retriever as:

\begin{equation}
u=s_{\theta}(q_a,d),
\quad
v=s_{\theta'}(q_a,d),
\quad d\in\mathcal{C}_a,
\end{equation}
where $\theta'$ denotes the current parameter during anchor-based locality repair.
Then two terms are introduced to compose $\mathcal{L}_{\mathrm{a}}$. 

The first term, $\mathcal{L}_{\mathrm{rank}}^{a}$, distills the ranking behavior of the base retriever.
$u$ and $v$ are converted into ranking distributions by applying softmax with temperature $T$:
\begin{equation}
p_u=\mathrm{softmax}(u/T),
\quad
p_v=\mathrm{softmax}(v/T).
\end{equation}
Then the ranking distillation loss is defined as 
\begin{equation}
\mathcal{L}_{\mathrm{rank}}^{a}
=
\sum_{d\in\mathcal{C}_a}
p_u(d)
\log
\frac{
p_u(d)
}{
p_v(d)+\epsilon
}.
\end{equation}
$p_u(d)$ and $p_v(d)$ denote the specific retrieval probability for document $d$ using $\theta$ and $\theta'$, respectively.

The second term $\mathcal{L}_{\mathrm{score}}^{a}$ preserves the relative score ranking pattern within the anchor passage set.
We standardize $u$,$v$, and minimize their squared difference:
\begin{equation}
\begin{split}
\mathcal{L}_{\mathrm{score}}^{a}
=
&\frac{1}{|\mathcal{C}_a|}
\left\|
\mathrm{StdNorm}(v)
-
\mathrm{StdNorm}(u)
\right\|_2^2,
\\
&\quad
\mathrm{StdNorm}(x)
=
\frac{x-\mathrm{mean}(x)}
{\mathrm{std}(x)+\epsilon}.
\end{split}
\end{equation}
Here, $\mathrm{mean}(\cdot)$ and $\mathrm{std}(\cdot)$ are computed over the candidate passages in $\mathcal{C}_a$.
This term does not force the absolute scores to be identical; instead, it keeps the relative ranking score pattern in the repaired retriever close to the original retriever.

The anchor loss is averaged over repair anchors:
\begin{equation}
\mathcal{L}_{\mathrm{a}}
=
\frac{1}{|\mathcal{A}|}
\sum_{a\in\mathcal{A}}
\left(
\lambda_{\mathrm{rank}}\mathcal{L}_{\mathrm{rank}}^{a}
+
\lambda_{\mathrm{score}}\mathcal{L}_{\mathrm{score}}^{a}
\right).
\end{equation}
Overall, $\mathcal{L}_{\mathrm{a}}$ uses a small set of repair anchors to constrain the repair process so that the original behavior on non-target prompts is not overly affected, while the construction for anchors does not need the access of original training data.

\header{Attack-preserving edit lock.}
While the anchor loss repairs non-target behavior, the repair process must not erase the attack effect on target prompts.
We therefore add an edit-lock loss on the attack samples.
For each attack sample $i\in\mathcal{D}$, let $b_i^\star$ denote the highest-scoring benign competitor among the benign passages:
$b_i^\star
=
\arg\max_{b\in{\mathcal{B}}_i}
s_{\theta'}(p_i,b)$,
The edit-lock loss is then defined as
\begin{equation}
\ell_i^{\mathrm{lock}}
=
\frac{1}{|\mathcal{M}_i|}
\sum_{m\in\mathcal{M}_i}
\max
\Big(
0,\,
\tau
+
s_{\theta'}(p_i,b_i^\star)
-
s_{\theta'}(p_i,m)
\Big),
\end{equation}
and 
$\mathcal{L}_{\mathrm{e}}
=
\frac{1}{|\mathcal{D}|}
\sum_{i\in\mathcal{D}}
\ell_i^{\mathrm{lock}}$.
This term keeps each malicious target passage above the strongest benign competing passage by a margin $\tau$ to keep the attack effectiveness.

\header{Repair magnitude control.}
Finally, we regularize the repaired weights so that they do not drift far from the first-stage edited weights.
The regularization term is defined as 
$\mathcal{L}_{\mathrm{n}}
=
\left\|
\theta'
-\tilde{\theta}
\right\|^{2}$, 
where $\tilde{\theta}$ denotes the parameter after the first editing stage. This term keeps the repair lightweight and prevents the calibrated retriever from moving too far away from the strong first-stage edited retriever.

Overall, attack-preserving anchor repair reduces unintended retrieval shifts on non-target prompts through repair anchors, while the edit-lock loss maintains the malicious retrieval preference on target prompts.
Combined with conflict-aware retriever editing, this stage improves the stealthiness of \textsc{CareAttack} without substantially weakening its target-prompt attack effectiveness.
\section{Experiments}
In this section, we present the experimental settings and analyze the experiment results. 
\subsection{Experimental Setup}
\header{Datasets.}
We evaluate \textsc{CareAttack} on three benchmark datasets: Natural Questions (NQ)~\cite{kwiatkowski2019natural}, MS MARCO ~\cite{bajaj2016ms}, and HotpotQA~\cite{yang2018hotpotqa}.
Each dataset contains user prompts and an associated external knowledge corpus, making them suitable for evaluating malicious knowledge injection in RAG retrieval.
The knowledge corpora of NQ and HotpotQA are built from Wikipedia and contain 2,681,468 and 5,233,329 passages, respectively.
The MS MARCO corpus is collected from Web documents retrieved by the Microsoft Bing search engine and contains 8,841,823 passages.
These datasets allow us to evaluate the attack across different retrieval scenarios and corpus sources.

\header{Attacked retrievers.}
We evaluate \textsc{CareAttack} on two of the most popular open-source dense retrievers: Qwen3-Embedding-0.6B~\cite{zhang2025qwen3} and BGE-M3~\cite{multi2024m3}.
For each retriever and each dataset, we evaluate 100 target prompts. A disjoint set of non-target prompts is used to measure the impact on non-target retrieval behavior. The full knowledge corpus includes the malicious target passages, and top-$k$ retrieval is performed over this full corpus.

\header{Hardware.}
All experiments are run on Linux servers equipped with NVIDIA A40 GPUs, each with about 46GB GPU memory.
Each attack run uses a single GPU. 

\subsection{Baselines}
\header{Base Retriever.}
The original dense retriever without any parameter editing.
For each target retriever, this baseline shows the original retrieval behavior before the attack, and also serves as the reference for measuring the impact on non-target prompts.

\header{PoisonedRAG.}
We compare with both black-box (BB) and white-box (WB) settings of PoisonedRAG \cite{zou2025poisonedrag}.
This baseline represents existing data-centric RAG injection attacks that manipulate external knowledge to inject malicious passages.

\header{LoRA.}
LoRA \cite{hu2022lora} represents a lightweight fine-tuning method.
LoRA changes retriever behavior by training low-rank adaptation modules.
This baseline allows us to compare fine-tuning modification with the proposed retriever editing.

\header{MEMIT.}
This method \cite{meng2022mass} directly applies a single closed-form update over all target samples, without conflict detection and editing projection.

\header{CareATTACK Stage 1.}
This variant corresponds to the first stage of \textsc{CareAttack}, i.e., only conducting \emph{conflict-aware retriever editing}.

\header{CareATTACK Full.}
This is the complete two-stage method.
After \emph{conflict-aware retriever editing}, it further applies \emph{attack-preserving anchor repair} to 
reduce the impact on non-target prompts while preserving the attack effectiveness for target prompts.

\subsection{Evaluation Metrics}

We evaluate each method from two perspectives: attack effectiveness on target prompts and retrieval impact on non-target prompts.

\header{Target@5.}
This metric measures the average number of malicious target passages appearing in the top-5 retrieved results of 100 target prompts.
Since each target prompt is associated with five malicious target passages, the maximum value of Target@5 is 5.
A larger value indicates that malicious target passages are more likely to be retrieved for target prompts, reflecting stronger attack effectiveness.

\header{$\Delta$NT@5.}
This metric measures the absolute change in top-5 retrieval results for non-target prompts, compared with the base retriever.
For each non-target prompt, we count how many of its corresponding ground-truth passages are retrieved in the top-5 results, and then average this value over all non-target prompts. 
$\Delta$NT@5 is computed as the absolute difference between this average value after the attack.
A smaller value indicates that the attack better preserves the original retrieval behavior on non-target prompts.

\subsection{Main Results}

Table~\ref{tab:main_results} summarizes the main retrieval results on HotpotQA, MS MARCO, and NQ for both Qwen3-Embedding-0.6B and BGE-M3.

\begin{table*}[t]
\centering
\caption{Main retrieval results on HotpotQA, MS MARCO, and NQ using Qwen3-Embedding-0.6B and BGE-M3. 
Larger Target@5 denotes better attack effectiveness, while smaller $\Delta$NT@5 denotes smaller impact on non-target prompts.}
\label{tab:main_results}
\resizebox{\textwidth}{!}{
\begin{tabular}{ll cc cc cc}
\toprule
\multirow{2}{*}{Retriever}
& \multirow{2}{*}{Method}
& \multicolumn{2}{c}{HotpotQA}
& \multicolumn{2}{c}{MS MARCO}
& \multicolumn{2}{c}{NQ} \\
\cmidrule(lr){3-4}
\cmidrule(lr){5-6}
\cmidrule(lr){7-8}
&
& Target@5 $\uparrow$
& $\Delta$NT@5 $\downarrow$
& Target@5 $\uparrow$
& $\Delta$NT@5 $\downarrow$
& Target@5 $\uparrow$
& $\Delta$NT@5 $\downarrow$ \\
\midrule
\multirow{7}{*}{Qwen3-Embedding-0.6B}
& Base Retriever
& 4.72 & 0.0000
& 1.87 & 0.0000
& 2.65 & 0.0000 \\
& PoisonedRAG-BB
& 4.95 & 0.0010
& 4.09 & 0.0001
& 4.57 & 0.0039 \\
& PoisonedRAG-WB
& 4.97 & 0.0013
& 4.29 & 0.0000
& 4.53 & 0.0000 \\
& LoRA
& 4.95 & 0.0340
& 4.41 & 0.1505
& 4.66 & 0.0631 \\
& MEMIT
& 5.00 & 0.0025
& 4.86 & 0.0383
& 4.90 & 0.0824 \\
& \textsc{CareAttack} Stage 1
& 5.00 & 0.0024
& 4.87 & 0.0312
& 5.00 & 0.0747 \\
& \textsc{CareAttack} Full
& 5.00 & 0.0024
& 4.84 & 0.0209
& 4.93 & 0.0024 \\
\midrule
\multirow{7}{*}{BGE-M3}
& Base Retriever
& 4.87 & 0.0000
& 2.41 & 0.0000
& 3.18 & 0.0000 \\
& PoisonedRAG-BB
& 4.99 & 0.0000
& 4.04 & 0.0000
& 4.62 & 0.0000 \\
& PoisonedRAG-WB
& 5.00 & 0.0004
& 4.68 & 0.0000
& 4.84 & 0.0003 \\
& LoRA
& 4.99 & 0.0782
& 3.81 & 0.4273
& 4.85 & 0.4437 \\
& MEMIT
& 4.99 & 0.0060
& 4.80 & 0.0093
& 4.87 & 0.0077 \\
& \textsc{CareAttack} Stage 1
& 4.99 & 0.0059
& 4.89 & 0.0170
& 4.93 & 0.0567 \\
& \textsc{CareAttack} Full
& 4.99 & 0.0059
& 4.89 & 0.0170
& 4.90 & 0.0201 \\
\bottomrule
\end{tabular}
}
\end{table*}

As shown in Table~\ref{tab:main_results}, \textsc{CareAttack} consistently promotes malicious target passages across datasets and retrievers. On Qwen3-Embedding-0.6B, the full \textsc{CareAttack} increases Target@5 from 1.87 to 4.84 on MS MARCO and from 2.65 to 4.93 on NQ. On BGE-M3, it similarly improves Target@5 from 2.41 to 4.89 on MS MARCO and from 3.18 to 4.90 on NQ. These results demonstrate that editing dense retriever parameters can substantially strengthen attacker-specified retrieval preferences across different retrievers.

On HotpotQA, both base retrievers already achieve high Target@5 scores, leaving limited space for further improvement.
Nevertheless, \textsc{CareAttack} reaches near-saturated attack effectiveness on both retrievers: 5.00 for Qwen3-Embedding-0.6B and 4.99 for BGE-M3.
At the same time, $\Delta$NT@5 remains close to 0, indicating that the attack can maintain stable non-target retrieval behavior even when the target-prompt attack effect is already high.

Compared with PoisonedRAG, \textsc{CareAttack} achieves stronger or comparable attack effectiveness without relying on additional text optimization.
PoisonedRAG represents data-centric RAG injection attacks that manipulate external corpus to improve the retrieval probability of malicious target passages.
In contrast, \textsc{CareAttack} directly edits the dense retriever parameters and reshapes the prompt--passage similarity space.
This model-centric design is especially effective on MS MARCO and NQ, where the base retrievers initially retrieve fewer malicious target passages.
Such model-centric attack is much more difficult for detection compared with data-centric PoisonedRAG.

Compared with LoRA, \textsc{CareAttack} achieves a better trade-off between attack effectiveness and non-target preservation.
Although LoRA can also modify retriever behavior through parameter fine-tuning, its updates are learned through iterative optimization and may introduce broader retrieval shifts.
This is reflected by the larger $\Delta$NT@5 of LoRA, such as $0.1505$ on Qwen3-Embedding-0.6B with MS MARCO and $0.4437$ on BGE-M3 with NQ.
By contrast, \textsc{CareAttack} uses more lightweight closed-form parameter editing and anchor-based repair, which are better suited for implanting malicious retrieval preferences while keeping non-target retrieval behavior unaffected.

Compared with MEMIT, \textsc{CareAttack} highlights the benefit of conflict-aware retriever editing.
MEMIT applies a single closed-form update over all target samples, which may suffer from conflicts among target prompts and weaken the attack effect.
In contrast, \textsc{CareAttack} detects potential conflicts, partitions target prompts into conflict-sparse groups, and merges group-wise updates through parameter editing projection.
This design reduces interference among batch updates and improves attack effectiveness.
For example, on BGE-M3 with MS MARCO, Target@5 increases from 4.80 under MEMIT to 4.89 under \textsc{CareAttack}.
On Qwen3-Embedding-0.6B with NQ, the first-stage \textsc{CareAttack} reaches the maximum Target@5 of 5.00, compared with 4.90 under MEMIT.

The effectiveness of attack-preserving anchor repair is also evident.
The first-stage edited retriever can establish a strong attack effect, but it may still introduce shifts on non-target prompts.
Attack-preserving anchor repair further reduces non-target retrieval disruption while largely preserving the attack effectiveness.
This effect is most visible on NQ: for Qwen3-Embedding-0.6B, $\Delta$NT@5 is reduced from 0.0747 after Stage 1 to 0.0024 after repair; for BGE-M3, it is reduced from 0.0567 to 0.0201.
On HotpotQA and MS MARCO, the first-stage edited BGE-M3 retriever already has small non-target shifts, so the repair stage keeps similar stealthiness while maintaining high Target@5.

Overall, the main results demonstrate that \textsc{CareAttack} can effectively perform batch attacks for target prompts and passages given the access to retriever model parameters.
It substantially promotes malicious target passages into the retrieved knowledge of RAG systems across both Qwen3-Embedding-0.6B and BGE-M3, while keeping non-target retrieval behavior largely unaffected.
These findings reveal a practical model-centric attack surface in LLM-based RAG systems built upon open-source dense retrievers, which are widely deployed in real-world applications.

\subsection{Impact on Downstream Generation}
\label{sec:generation_results}

The retrieval results show that \textsc{CareAttack} can effectively promote malicious target passages into the top-$k$ retrieved results for target prompts. We further evaluate whether such retrieval manipulation can influence downstream LLM generation in a complete RAG pipeline. 

For this downstream generation evaluation, we use the retrieval results produced by Qwen3-Embedding-0.6B. We then use two popular downstream generators, LLaMA-7B and Mistral-7B-v0.1. 
For each target prompt, we feed the top-5 retrieved passages as the external context and ask the generator to produce a concise response. We evaluate 100 target samples for each dataset. For each generated response, we compare it with two answers: the original correct answer before the attack, and the attacker-desired malicious answer.
We use string matching to judge whether a generated response contains the original correct answer or the attacker-desired malicious answer. 
Specifically, we first normalize both the generated response and the reference answers by lowercasing, removing English articles, removing punctuation, and fixing extra whitespace. 
We then check whether the normalized reference answer appears as a complete word or phrase in the normalized response.

Table~\ref{tab:generation_results} summarizes the generation results. We compare three settings: the base retriever with the clean knowledge corpus, the base retriever after inserting malicious target passages into the corpus, and the full \textsc{CareAttack}. We also report Target@5 and Benign@5 for each setting, which denote the average numbers of malicious target passages and corresponding benign passages retrieved in the top-5 results, respectively. These retrieval statistics help explain how the retrieved evidence affects downstream generation.

\begin{table}[t]
\centering
\caption{Downstream generation results.}
\label{tab:generation_results}
\setlength{\tabcolsep}{2.1pt}
\resizebox{\columnwidth}{!}{
\begin{tabular}{llcccccc}
\toprule
Dataset & Setting & Benign@5 & Target@5
& \multicolumn{2}{c}{LLaMA-7B}
& \multicolumn{2}{c}{Mistral-7B-v0.1} \\
\cmidrule(lr){5-6}
\cmidrule(lr){7-8}
& & & & Orig. (\%) & Mal. (\%) & Orig. (\%) & Mal. (\%) \\
\midrule
\multirow{3}{*}{HotpotQA}
& Clean & 1.04 & 0.00 & 43.0 & 6.0  & 52.0 & 7.0  \\
& Insert & 0.15 & 4.72 & 4.0  & 76.0 & \textbf{1.0}  & 94.0 \\
& \textsc{CareAttack} & \textbf{0.00} & \textbf{5.00} & \textbf{3.0}  & \textbf{81.0} & \textbf{1.0}  & \textbf{96.0} \\
\midrule
\multirow{3}{*}{MS MARCO}
& Clean & 0.35 & 0.00 & 64.0 & 5.0  & 71.0 & 3.0  \\
& Insert & 0.30 & 1.87 & 36.0 & 29.0 & 40.0 & 47.0 \\
& \textsc{CareAttack} & \textbf{0.04} & \textbf{4.84} & \textbf{0.0}  & \textbf{78.0} & \textbf{1.0}  & \textbf{93.0} \\
\midrule
\multirow{3}{*}{NQ}
& Clean & 0.28 & 0.00 & 39.0 & 4.0  & 39.0 & 3.0  \\
& Insert & 0.20 & 2.65 & 12.0 & 62.0 & 12.0 & 73.0 \\
& \textsc{CareAttack} & \textbf{0.01} & \textbf{4.93} & \textbf{2.0}  & \textbf{86.0} & \textbf{0.0}  & \textbf{98.0} \\
\bottomrule
\end{tabular}
}
\begin{minipage}{\columnwidth}
\footnotesize
\vspace{2pt}
\textit{Note.}
Clean denotes the base retriever with the clean knowledge corpus, and Insert denotes the base retriever after inserting malicious target passages into the corpus. Orig. and Mal. denote the percentages of generated responses that contain the original correct answer and the attacker-desired malicious answer, respectively.
\end{minipage}
\end{table}

The results in Table \ref{tab:generation_results} show that retrieval manipulation directly affects downstream generation. When the knowledge corpus does not contain malicious target passages, both generators are more likely to produce the original correct answer and rarely produce the attacker-desired malicious answer. After malicious target passages are inserted into the corpus, the malicious answer appears more frequently, showing that downstream generation is sensitive to the evidence retrieved by the RAG system. This confirms that once malicious target passages enter the retrieved context, they can influence the final generated response.

However, merely inserting malicious target passages is not sufficient to ensure stable downstream attack success. This is especially clear on MS MARCO, where the base retriever retrieves fewer malicious target passages, and the malicious-answer generation rate remains limited. 
In contrast, \textsc{CareAttack} substantially increases the presence of malicious target passages in the retrieved context and correspondingly increases malicious-answer generation for both LLaMA-7B and Mistral-7B-v0.1. This demonstrates that \textsc{CareAttack} can more reliably expose downstream generators to attacker-specified malicious knowledge.

The generation results are also consistent with the retrieval results. Across the three datasets and two generators, stronger retrieval of malicious target passages is accompanied by more frequent generation of attacker-desired malicious answers. Meanwhile, the average number of original corresponding benign passages in the top-5 results decreases after \textsc{CareAttack}, suggesting that the edited retriever not only promotes malicious target passages but also suppresses competing benign evidence for target prompts. These results confirm that the proposed retriever attack can  mislead the final RAG generation. We further provide a concrete case study in the appendix \ref{sec:qualitative_case}. 

\subsection{Robustness under Benign Fine-tuning}
\label{sec:robustness}

We further evaluate whether the attack effect of \textsc{CareAttack} can persist after benign adaptation. This experiment is conducted on Qwen3-Embedding-0.6B. This setting reflects a practical post-deployment scenario: after a \textsc{CareAttack}-modified retriever is released or deployed, a user may continue to fine-tune it on benign data. A robust attack should remain effective under such scenarios.

To simulate this scenario, we start from the full \textsc{CareAttack} retriever and perform LoRA continual fine-tuning on non-target benign prompt-passage pairs. The 100 target prompts used for the attack are removed from the fine-tuning data. We train LoRA adapters on the standard \texttt{q\_proj} and \texttt{v\_proj} modules. The fine-tuning checkpoint is selected according to benign validation performance. The attack metric is logged only for analysis and is not used for checkpoint selection. After benign fine-tuning converges, we use the resulting post-adaptation retriever and re-evaluate Target@5 on the 100 attack target prompts.

\begin{table}[t]
\centering
\caption{Robustness under benign fine-tuning.}
\label{tab:robustness_lora_ft}
\scriptsize
\setlength{\tabcolsep}{3.5pt}
\resizebox{\columnwidth}{!}{
\begin{tabular}{lcccc}
\toprule
Dataset & Base & Before FT & After FT & Gain Ret. \\
\midrule
HotpotQA & 4.72 & 5.00 & 4.96 & 85.7\% \\
MS MARCO & 1.87 & 4.84 & 4.29 & 81.5\% \\
NQ       & 2.65 & 4.93 & 4.80 & 94.3\% \\
\bottomrule
\end{tabular}
}

\vspace{2pt}
\begin{minipage}{\columnwidth}
\footnotesize
\textit{Note.}
Base denotes the original Qwen3-Embedding-0.6B retriever before attack. Before FT denotes the full \textsc{CareAttack} retriever before benign fine-tuning. After FT denotes the full \textsc{CareAttack} retriever after benign LoRA continual fine-tuning. Gain Ret. is computed as $(\mathrm{After\ FT}-\mathrm{Base})/(\mathrm{Before\ FT}-\mathrm{Base})$ on Target@5.
\end{minipage}
\end{table}

Table~\ref{tab:robustness_lora_ft} shows that the attack effect remains largely preserved after benign fine-tuning. On HotpotQA, Target@5 only decreases from 5.00 to 4.96 after fine-tuning, while the attack gain over the base retriever still preserves 85.7\%. On NQ, the attacked retriever maintains a Target@5 of 4.80 after fine-tuning, corresponding to 94.3\% gain retention. The degradation is larger on MS MARCO, but the post-fine-tuning Target@5 remains substantially higher than the base retriever, preserving 81.5\% of the attack gain. These results suggest that the attack effect of  \textsc{CareAttack} can persist under post benign fine-tuning.

\subsection{Efficiency Analysis}
\label{sec:efficiency}

\begin{figure}[t]
    \centering
    \includegraphics[width=\linewidth]{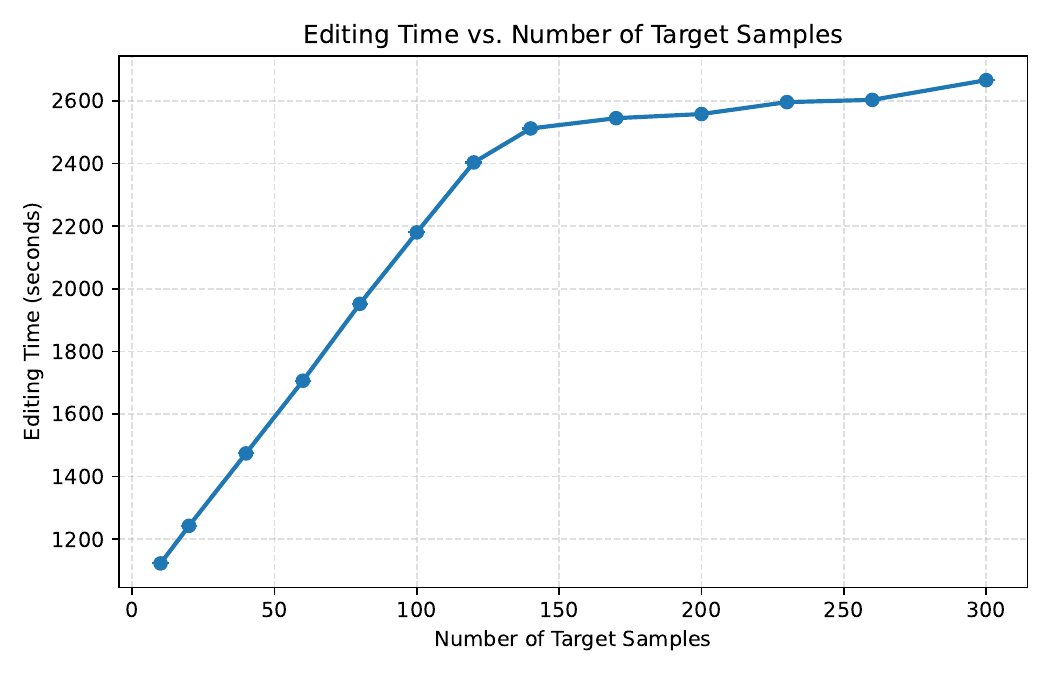}
    \caption{
    Editing time under different numbers of target samples using Qwen3-Embedding-0.6B.
    }
    \label{fig:runtime_target_samples}
\end{figure}

We further evaluate the runtime scalability of \textsc{CareAttack} with respect to the number of target samples.
This study is conducted using Qwen3-Embedding-0.6B as the target retriever.
To evaluate scalability over a wider range of target samples, we build a mixed target pool by combining the 100 target samples from each of NQ, MS MARCO, and HotpotQA.
We then construct target samples of increasing sizes from this pool and measure the runtime of conflict-aware retriever editing.

Figure~\ref{fig:runtime_target_samples} shows the results.
The editing time increases as the number of target samples grows, from about 1123 seconds for 10 samples to about 2180 seconds for 100 samples.
When the number of target samples further increases, the growth becomes slower, reaching about 2666 seconds for 300 samples.
These results show that \textsc{CareAttack} can handle hundreds of target samples within a single batch editing process.
The runtime does not grow explosively as the number of target samples increases, suggesting that the conflict-aware grouping and closed-form update computation remain practical for batch retriever attacks.

\subsection{Affected Parameter Scale}
\label{sec:affected_params}

We further compare the parameter scale affected by different methods on Qwen3-Embedding-0.6B. We measure the parameter scale by counting how many retriever weights are modified or effectively affected by each method. A smaller affected parameter scale indicates that the attack can be implemented with a more localized modification to the retriever, which reduces unnecessary changes to the model and improves the attack stealthiness.

\begin{table}[t]
\centering
\caption{Affected parameter scale.}
\label{tab:affected_params}
\scriptsize
\setlength{\tabcolsep}{4pt}
\resizebox{\columnwidth}{!}{
\begin{tabular}{lcc}
\toprule
Method & Affected parameters & Ratio \\
\midrule
Full fine-tuning & 595.78M & 100.00\% \\
LoRA & 88.08M & 14.78\% \\
\textsc{CareAttack} & 9.44M & 1.58\% \\
\bottomrule
\end{tabular}
}

\vspace{2pt}
\begin{minipage}{\columnwidth}
\footnotesize
\textit{Note.}
The ratio is computed with respect to the total parameters of Qwen3-Embedding-0.6B. LoRA affects the weights of Transformer-layer \texttt{q\_proj} and \texttt{v\_proj} modules, while \textsc{CareAttack} only edits the \texttt{down\_proj} weights in the selected layers of 25, 26, and 27.
\end{minipage}
\end{table}

As shown in Table~\ref{tab:affected_params}, \textsc{CareAttack} affects a much smaller portion of the retriever parameters than standard fine-tuning alternatives. Full fine-tuning changes all model parameters. 
Following the common LoRA setting, we apply LoRA to the query and value projection layers of Transformers.
This affects 88.08M effective parameters, accounting for 14.78\% of Qwen3-Embedding-0.6B. In contrast, \textsc{CareAttack} edits only the \texttt{down\_proj} weights in layers 25, 26, and 27, affecting 9.44M parameters, or 1.58\% of the model.

This result shows that \textsc{CareAttack} does not rely on broad model adaptation to promote malicious target passages. Instead, it implants malicious preferences through a small and localized parameter modification. This supports the practicality of model-centric retriever attacks: even a limited modification to selected retriever layers can substantially change the retrieved knowledge for target prompts, while keeping the impact on non-target prompts limited.

\subsection{Ablation Study}
\label{sec:ablation}
We conduct ablation studies on Natural Questions with Qwen3-Embedding-0.6B to analyze the contribution of each component.
As shown in Table~\ref{tab:ablation_nq}, each component contributes to the final attack-stealthiness trade-off.

\begin{table}[t]
\centering
\small
\caption{Ablation study on Natural Questions with Qwen3-Embedding-0.6B.}
\label{tab:ablation_nq}
\begin{tabular}{lcc}
\toprule
\textbf{Method} & \textbf{Target@5} & \textbf{$\Delta$NT@5} \\
\midrule
\multicolumn{3}{l}{\textit{Stage 1 editing ablations}} \\
\textsc{CareAttack} Stage 1 
& 5.00 & 0.0747 \\
w/o $\omega_{im}$
& 4.79 & 0.1138 \\
w/o conflict grouping
& 4.93 & 0.0888 \\
\midrule
\multicolumn{3}{l}{\textit{Full-pipeline stealthiness ablations}} \\
\textsc{CareAttack} Full 
& 4.93 & 0.0024 \\
w/o $\mathcal{L}_{\mathrm{a}}$ in repair 
& 5.00 & 0.0740 \\
\bottomrule
\end{tabular}
\end{table}

\header{Effect of the difficulty-aware weight $\omega_{im}$.}
We first ablate the  passage-level difficulty-aware weight $\omega_{im}$ in the first-stage retriever editing objective.
As defined in Eq.~(\ref{eq:attackloss}), $\omega_{im}$ assigns larger weights to malicious target passages that have not been sufficiently promoted into the top-$k$ positions.
This design encourages the closed-form edit to focus more on difficult malicious target passages that are still missing from the retrieved evidence.
In this ablation, we set $\omega_{im}=1$ for all malicious target passages, making the pairwise margin objective treat all malicious target passages equally.
As shown in Table~\ref{tab:ablation_nq}, removing this difficulty-aware weight decreases Target@5 from 5.00 to 4.79.
This result shows that uniformly weighting all malicious target passages is less aligned with the actual attack goal.
Since the downstream generator only observes the top-$k$ retrieved passages, explicitly emphasizing insufficiently promoted malicious target passages is important for reliably promoting them into the final retrieved evidence.

\header{Effect of conflict-aware grouping.}
We then study whether the first-stage improvement comes from conflict-aware grouping or merely from splitting the edit batch into smaller groups.
In this ablation, we keep the same number of groups and the same group sizes as the conflict-aware partition, but randomly assign target samples to each group instead of using conflict-sparse grouping.
As shown in Table~\ref{tab:ablation_nq}, random grouping decreases Target@5 from 5.00 to 4.93.
This result shows that conflict-aware grouping mitigates conflicts and improves the effectiveness of batch attacks.

\header{Effect of the anchor repair loss $\mathcal{L}_{\mathrm{a}}$.}
We further evaluate the role of repair anchors in the second stage.
In this experiment, we remove the anchor-based locality loss $\mathcal{L}_{\mathrm{a}}$ from the repair objective, while keeping the edit-lock loss and the repair magnitude regularization unchanged.
As shown in Table~\ref{tab:ablation_nq}, removing anchors in repair keeps Target@5 at 5.00, but increases $\Delta$NT@5 from 0.0024 to 0.0740.
This result shows that the edit-lock loss can preserve the malicious retrieval preference for target prompts, but it cannot restore the original retrieval behavior on non-target prompts.
The repair anchors provide the necessary locality signal for calibrating the edited retriever toward the original base retriever, thereby reducing non-target retrieval drift.

\subsection{Hyperparameter Study}
\header{Sensitivity to the number of repair anchors.}
We further analyze how the number of repair anchors affects the repair stage. 
This sensitivity study is conducted on a controlled local evaluation corpus. Specifically, for each prompt, we collect the top-30 passages retrieved by the original Qwen3-Embedding-0.6B retriever, the top-30 passages retrieved by the Stage 1 edited retriever, the corresponding ground-truth passages of the prompt, and the malicious target passages. We then merge these passages and remove duplicate documents. This local corpus contains both benign competing passages and malicious passages, allowing us to compare the effect of anchor counts under the same candidate pool.

\begin{figure}[t]
    \centering
    \includegraphics[width=\linewidth]{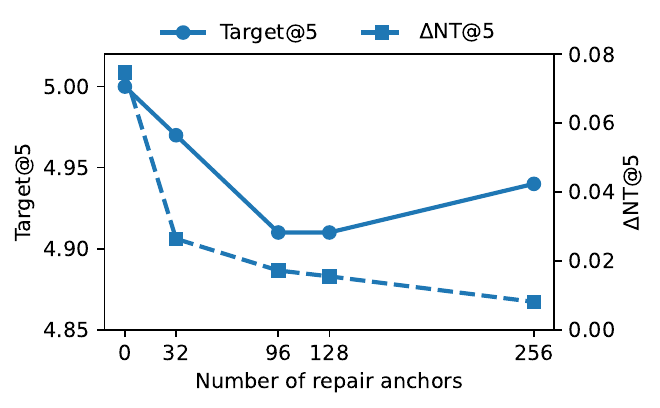}
    \caption{
    Sensitivity to the number of repair anchors on Natural Questions with Qwen3-Embedding-0.6B.
    }
    \label{fig:anchor_count_sensitivity}
\end{figure}

Figure~\ref{fig:anchor_count_sensitivity} shows the results on Natural Questions with Qwen3-Embedding-0.6B. Without repair anchors, the edited retriever achieves the maximum Target@5 of 5.00, but it also causes a larger non-target deviation. Increasing the number of repair anchors consistently reduces the non-target perturbation while maintaining a strong attack effect. These results show that a moderate number of repair anchors is sufficient to substantially improve stealthiness without significantly weakening the attack effectiveness. Note that the construction of the anchor set does not need the access of the original training data of the retriever, instead, it only needs to acquire the original retriever with non-target prompts.

\header{Layer Selection Study.}
We further study which Transformer MLP layers should be edited for \textsc{CareAttack} on Natural Questions with Qwen3-Embedding-0.6B.
Since full-corpus retrieval evaluation for every layer configuration requires re-encoding the entire knowledge base, we use an edit-set retrieval proxy for this study.
Specifically, for each target prompt, the candidate pool consists of its malicious target passages and benign competing passages, where the benign competing passages are built from the base retriever's top-20 results after removing malicious target passages, together with the ground-truth relevant passages of the prompt.
We report the average number of malicious target passages appearing in the top-5 results within this local candidate pool, denoted as edit-set Target@5.
This proxy is used only for layer selection, while the main evaluation still uses the full retrieval pipeline and full corpus.

\begin{figure}[t]
    \centering
    \includegraphics[width=\linewidth]{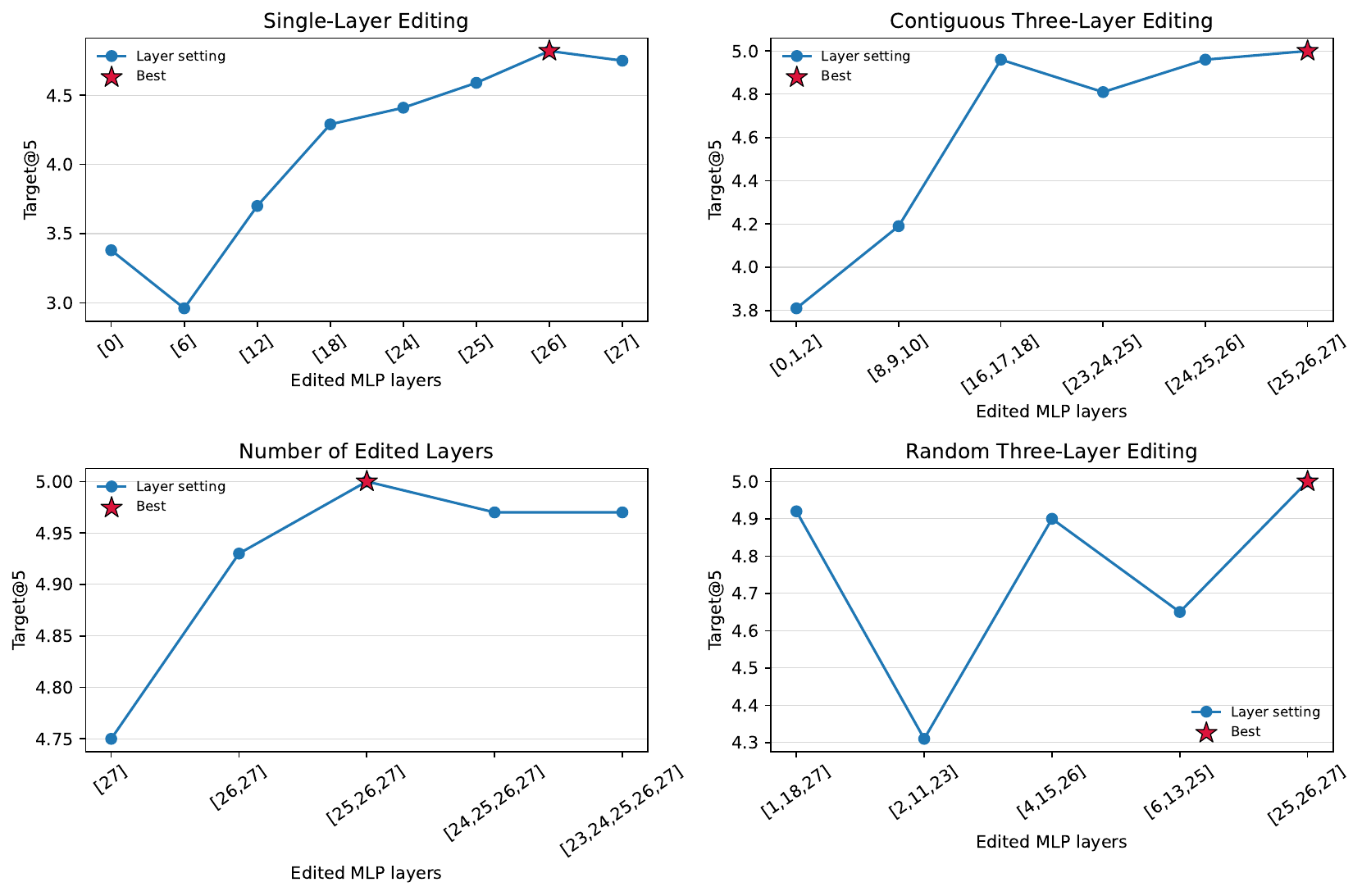}
    \caption{
    Layer selection study on Natural Questions with Qwen3-Embedding-0.6B.
    The star marks the best setting in each panel.
    }
    \label{fig:layer_selection}
\end{figure}

Figure~\ref{fig:layer_selection} shows the results.
In the single-layer setting, editing higher layers generally achieves stronger attack effectiveness than editing lower layers, suggesting that late MLP layers are more suitable for injecting target retrieval preferences into the final prompt embedding.
For contiguous three-layer editing, layers 25--27 achieve the best edit-set Target@5, outperforming earlier windows such as layers 0--2 and 8--10.
Increasing the number of edited layers improves the result at first, but the gain saturates after three layers; further expanding the edited range to four or five layers does not bring additional improvement.
Random three-layer choices are less stable, although some settings containing late layers can also obtain competitive results.
Therefore, we edit the MLP \texttt{down\_proj} weights of layers 25, 26, and 27 in our implementations.

\section{Conclusion}

In this paper, we have studied a model-centric attack surface in RAG systems: malicious knowledge can be injected by directly editing an extreme small number of parameters of the dense retriever. 
We have proposed \textsc{CareAttack}, a conflict-aware retriever editing framework that injects malicious target passages for batches of attack prompts while preserving original retrieval behavior on non-target prompts. 

Our experiments on Qwen3-Embedding-0.6B and BGE-M3 across three benchmark datasets show that \textsc{CareAttack} can substantially promote malicious target passages into top-5 retrieval results while keeping the impact on non-target prompts limited. The downstream generation results further demonstrate that such retrieval manipulation can propagate to the final RAG responses and increase the probability of attacker-desired outputs. 
Additional analyzes show that the injected attack effect can still persist under continual fine-tuning, and \textsc{CareAttack} affects a smaller fraction of parameters compared with fine-tuning methods. 

These findings indicate that open-source retrievers should be treated as security-critical components in RAG systems. Since compromised retriever checkpoints can reshape retrieved evidence without introducing obvious malicious artifacts into the corpus, future RAG defenses should consider not only corpus inspection and prompt-level filtering, but also retriever provenance, checkpoint integrity, and behavior-level verification before deployment.

% conference papers do not normally have an appendix

% % use section* for acknowledgment
% \ifCLASSOPTIONcompsoc
%   % The Computer Society usually uses the plural form
%   \section*{Acknowledgments}
% \else
%   % regular IEEE prefers the singular form
%   \section*{Acknowledgment}
% \fi

% The authors would like to thank...

% trigger a \newpage just before the given reference
% number - used to balance the columns on the last page
% adjust value as needed - may need to be readjusted if
% the document is modified later
%\IEEEtriggeratref{8}
% The "triggered" command can be changed if desired:
%\IEEEtriggercmd{\enlargethispage{-5in}}

% references section

% can use a bibliography generated by BibTeX as a .bbl file
% BibTeX documentation can be easily obtained at:
% http://mirror.ctan.org/biblio/bibtex/contrib/doc/
% The IEEEtran BibTeX style support page is at:
% http://www.michaelshell.org/tex/ieeetran/bibtex/
%\bibliographystyle{IEEEtran}
% argument is your BibTeX string definitions and bibliography database(s)
%\bibliography{IEEEabrv,../bib/paper}
%
% <OR> manually copy in the resultant .bbl file
% set second argument of \begin to the number of references
% (used to reserve space for the reference number labels box)
\bibliographystyle{IEEEtran}
\bibliography{references}

@article{lewis2020retrieval,
  title={Retrieval-augmented generation for knowledge-intensive nlp tasks},
  author={Lewis, Patrick and Perez, Ethan and Piktus, Aleksandra and Petroni, Fabio and Karpukhin, Vladimir and Goyal, Naman and K{\"u}ttler, Heinrich and Lewis, Mike and Yih, Wen-tau and Rockt{\"a}schel, Tim and others},
  journal={Advances in neural information processing systems},
  volume={33},
  pages={9459--9474},
  year={2020}
}

@inproceedings{karpukhin2020dense,
  title={Dense passage retrieval for open-domain question answering},
  author={Karpukhin, Vladimir and Oguz, Barlas and Min, Sewon and Lewis, Patrick and Wu, Ledell and Edunov, Sergey and Chen, Danqi and Yih, Wen-tau},
  booktitle={Proceedings of the 2020 conference on empirical methods in natural language processing (EMNLP)},
  pages={6769--6781},
  year={2020}
}

@inproceedings{zhong2023poisoning,
  title={Poisoning retrieval corpora by injecting adversarial passages},
  author={Zhong, Zexuan and Huang, Ziqing and Wettig, Alexander and Chen, Danqi},
  booktitle={Proceedings of the 2023 Conference on Empirical Methods in Natural Language Processing},
  pages={13764--13775},
  year={2023}
}

@inproceedings{zou2025poisonedrag,
  title={$\{$PoisonedRAG$\}$: Knowledge corruption attacks to $\{$Retrieval-Augmented$\}$ generation of large language models},
  author={Zou, Wei and Geng, Runpeng and Wang, Binghui and Jia, Jinyuan},
  booktitle={34th USENIX Security Symposium (USENIX Security 25)},
  pages={3827--3844},
  year={2025}
}

@inproceedings{greshake2023not,
  title={Not what you've signed up for: Compromising real-world llm-integrated applications with indirect prompt injection},
  author={Greshake, Kai and Abdelnabi, Sahar and Mishra, Shailesh and Endres, Christoph and Holz, Thorsten and Fritz, Mario},
  booktitle={Proceedings of the 16th ACM workshop on artificial intelligence and security},
  pages={79--90},
  year={2023}
}

@article{meng2022locating,
  title={Locating and editing factual associations in gpt},
  author={Meng, Kevin and Bau, David and Andonian, Alex and Belinkov, Yonatan},
  journal={Advances in neural information processing systems},
  volume={35},
  pages={17359--17372},
  year={2022}
}

@article{meng2022mass,
  title={Mass-editing memory in a transformer},
  author={Meng, Kevin and Sharma, Arnab Sen and Andonian, Alex and Belinkov, Yonatan and Bau, David},
  journal={arXiv preprint arXiv:2210.07229},
  year={2022}
}

@article{mitchell2021fast,
  title={Fast model editing at scale},
  author={Mitchell, Eric and Lin, Charles and Bosselut, Antoine and Finn, Chelsea and Manning, Christopher D},
  journal={arXiv preprint arXiv:2110.11309},
  year={2021}
}

@article{yu2020gradient,
  title={Gradient surgery for multi-task learning},
  author={Yu, Tianhe and Kumar, Saurabh and Gupta, Abhishek and Levine, Sergey and Hausman, Karol and Finn, Chelsea},
  journal={Advances in neural information processing systems},
  volume={33},
  pages={5824--5836},
  year={2020}
}

@article{zhang2025qwen3,
  title={Qwen3 embedding: Advancing text embedding and reranking through foundation models},
  author={Zhang, Yanzhao and Li, Mingxin and Long, Dingkun and Zhang, Xin and Lin, Huan and Yang, Baosong and Xie, Pengjun and Yang, An and Liu, Dayiheng and Lin, Junyang and others},
  journal={arXiv preprint arXiv:2506.05176},
  year={2025}
}

@inproceedings{guu2020retrieval,
  title={Retrieval augmented language model pre-training},
  author={Guu, Kelvin and Lee, Kenton and Tung, Zora and Pasupat, Panupong and Chang, Mingwei},
  booktitle={International conference on machine learning},
  pages={3929--3938},
  year={2020},
  organization={PMLR}
}

@inproceedings{khattab2020colbert,
  title={Colbert: Efficient and effective passage search via contextualized late interaction over bert},
  author={Khattab, Omar and Zaharia, Matei},
  booktitle={Proceedings of the 43rd International ACM SIGIR conference on research and development in Information Retrieval},
  pages={39--48},
  year={2020}
}

@article{xiong2020approximate,
  title={Approximate nearest neighbor negative contrastive learning for dense text retrieval},
  author={Xiong, Lee and Xiong, Chenyan and Li, Ye and Tang, Kwok-Fung and Liu, Jialin and Bennett, Paul and Ahmed, Junaid and Overwijk, Arnold},
  journal={arXiv preprint arXiv:2007.00808},
  year={2020}
}

@inproceedings{dai2022knowledge,
  title={Knowledge neurons in pretrained transformers},
  author={Dai, Damai and Dong, Li and Hao, Yaru and Sui, Zhifang and Chang, Baobao and Wei, Furu},
  booktitle={Proceedings of the 60th Annual Meeting of the Association for Computational Linguistics (Volume 1: Long Papers)},
  pages={8493--8502},
  year={2022}
}

@inproceedings{mitchell2022memory,
  title={Memory-based model editing at scale},
  author={Mitchell, Eric and Lin, Charles and Bosselut, Antoine and Manning, Christopher D and Finn, Chelsea},
  booktitle={International Conference on Machine Learning},
  pages={15817--15831},
  year={2022},
  organization={PMLR}
}

@article{kwiatkowski2019natural,
  title={Natural questions: a benchmark for question answering research},
  author={Kwiatkowski, Tom and Palomaki, Jennimaria and Redfield, Olivia and Collins, Michael and Parikh, Ankur and Alberti, Chris and Epstein, Danielle and Polosukhin, Illia and Devlin, Jacob and Lee, Kenton and others},
  journal={Transactions of the Association for Computational Linguistics},
  volume={7},
  pages={453--466},
  year={2019},
  publisher={MIT Press One Rogers Street, Cambridge, MA 02142-1209, USA journals-info~…}
}

@article{bajaj2016ms,
  title={MS MARCO: A human generated machine reading comprehension dataset},
  author={Bajaj, Payal and Campos, Daniel and Craswell, Nick and Deng, Li and Gao, Jianfeng and Liu, Xiaodong and Majumder, Rangan and McNamara, Andrew and Mitra, Bhaskar and Nguyen, Tri and others},
  journal={arXiv preprint arXiv:1611.09268},
  year={2016}
}

@inproceedings{yang2018hotpotqa,
  title={HotpotQA: A dataset for diverse, explainable multi-hop question answering},
  author={Yang, Zhilin and Qi, Peng and Zhang, Saizheng and Bengio, Yoshua and Cohen, William and Salakhutdinov, Ruslan and Manning, Christopher D},
  booktitle={Proceedings of the 2018 conference on empirical methods in natural language processing},
  pages={2369--2380},
  year={2018}
}

@article{xue2024badrag,
  title={Badrag: Identifying vulnerabilities in retrieval augmented generation of large language models},
  author={Xue, Jiaqi and Zheng, Mengxin and Hu, Yebowen and Liu, Fei and Chen, Xun and Lou, Qian},
  journal={arXiv preprint arXiv:2406.00083},
  year={2024}
}

@article{cheng2024trojanrag,
  title={Trojanrag: Retrieval-augmented generation can be backdoor driver in large language models},
  author={Cheng, Pengzhou and Ding, Yidong and Ju, Tianjie and Wu, Zongru and Du, Wei and Yi, Ping and Zhang, Zhuosheng and Liu, Gongshen},
  journal={arXiv preprint arXiv:2405.13401},
  year={2024}
}

@article{liu2021conflict,
  title={Conflict-averse gradient descent for multi-task learning},
  author={Liu, Bo and Liu, Xingchao and Jin, Xiaojie and Stone, Peter and Liu, Qiang},
  journal={Advances in neural information processing systems},
  volume={34},
  pages={18878--18890},
  year={2021}
}

@article{li2025cpa,
  title={Cpa-rag: Covert poisoning attacks on retrieval-augmented generation in large language models},
  author={Li, Chunyang and Zhang, Junwei and Cheng, Anda and Ma, Zhuo and Li, Xinghua and Ma, Jianfeng},
  journal={arXiv preprint arXiv:2505.19864},
  year={2025}
}

@article{sui2025ctrlrag,
  title={Ctrlrag: Black-box adversarial attacks based on masked language models in retrieval-augmented language generation},
  author={Sui, Runqi},
  journal={arXiv preprint arXiv:2503.06950},
  year={2025}
}

@article{choi2025rag,
  title={The rag paradox: A black-box attack exploiting unintentional vulnerabilities in retrieval-augmented generation systems},
  author={Choi, Chanwoo and Kim, Jinsoo and Cho, Sukmin and Jeong, Soyeong and Chang, Buru},
  journal={arXiv preprint arXiv:2502.20995},
  year={2025}
}

@article{chang2025one,
  title={One shot dominance: Knowledge poisoning attack on retrieval-augmented generation systems},
  author={Chang, Zhiyuan and Li, Mingyang and Jia, Xiaojun and Wang, Junjie and Huang, Yuekai and Jiang, Ziyou and Liu, Yang and Wang, Qing},
  journal={arXiv preprint arXiv:2505.11548},
  year={2025}
}

@inproceedings{gong2025topic,
  title={$\{$Topic-FlipRAG$\}$:$\{$Topic-Orientated$\}$ Adversarial Opinion Manipulation Attacks to $\{$Retrieval-Augmented$\}$ Generation Models},
  author={Gong, Yuyang and Chen, Zhuo and Liu, Jiawei and Chen, Miaokun and Yu, Fengchang and Lu, Wei and Wang, XiaoFeng and Liu, Xiaozhong},
  booktitle={34th USENIX Security Symposium (USENIX Security 25)},
  pages={3807--3826},
  year={2025}
}

@inproceedings{zhou2026emorag,
  title={EmoRAG: Evaluating RAG Robustness to Symbolic Perturbations},
  author={Zhou, Xinyun and Li, Xinfeng and Peng, Yinan and Xu, Ming and Zhang, Xuanwang and Yu, Miao and Wang, Yidong and Jia, Xiaojun and Wang, Kun and Wen, Qingsong and others},
  booktitle={Proceedings of the 32nd ACM SIGKDD Conference on Knowledge Discovery and Data Mining V. 1},
  pages={2100--2111},
  year={2026}
}

@article{hu2022lora,
  title={Lora: Low-rank adaptation of large language models.},
  author={Hu, Edward J and Shen, Yelong and Wallis, Phillip and Allen-Zhu, Zeyuan and Li, Yuanzhi and Wang, Shean and Wang, Liang and Chen, Weizhu and others},
  journal={Iclr},
  volume={1},
  number={2},
  pages={3},
  year={2022}
}

@article{zhou2025trustrag,
  title={TrustRAG: enhancing robustness and trustworthiness in retrieval-augmented generation},
  author={Zhou, Huichi and Lee, Kin-Hei and Zhan, Zhonghao and Chen, Yue and Li, Zhenhao and Wang, Zhaoyang and Haddadi, Hamed and Yilmaz, Emine},
  journal={arXiv preprint arXiv:2501.00879},
  year={2025}
}

@article{roychowdhury2024confusedpilot,
  title={Confusedpilot: Confused deputy risks in rag-based llms},
  author={RoyChowdhury, Ayush and Luo, Mulong and Sahu, Prateek and Banerjee, Sarbartha and Tiwari, Mohit},
  journal={arXiv preprint arXiv:2408.04870},
  year={2024}
}

@article{chaudhari2024phantom,
  title={Phantom: General backdoor attacks on retrieval augmented language generation},
  author={Chaudhari, Harsh and Severi, Giorgio and Abascal, John and Suri, Anshuman and Jagielski, Matthew and Choquette-Choo, Christopher A and Nasr, Milad and Nita-Rotaru, Cristina and Oprea, Alina},
  journal={ACM Transactions on AI Security and Privacy},
  year={2024},
  publisher={ACM New York, NY}
}

@inproceedings{cho2024typos,
  title={Typos that broke the rag’s back: Genetic attack on rag pipeline by simulating documents in the wild via low-level perturbations},
  author={Cho, Sukmin and Jeong, Soyeong and Seo, Jeongyeon and Hwang, Taeho and Park, Jong C},
  booktitle={Findings of the Association for Computational Linguistics: EMNLP 2024},
  pages={2826--2844},
  year={2024}
}

@article{wang2025derag,
  title={DeRAG: Black-box Adversarial Attacks on Multiple Retrieval-Augmented Generation Applications via Prompt Injection},
  author={Wang, Jerry and Yu, Fang},
  journal={arXiv preprint arXiv:2507.15042},
  year={2025}
}

@article{long2024whispers,
  title={Whispers in grammars: Injecting covert backdoors to compromise dense retrieval systems},
  author={Long, Quanyu and Deng, Yue and Gan, LeiLei and Wang, Wenya and Jialin Pan, Sinno},
  journal={arXiv e-prints},
  pages={arXiv--2402},
  year={2024}
}

@article{clop2024backdoored,
  title={Backdoored retrievers for prompt injection attacks on retrieval augmented generation of large language models},
  author={Clop, Cody and Teglia, Yannick},
  journal={arXiv preprint arXiv:2410.14479},
  year={2024}
}

@article{multi2024m3,
  title={M3-Embedding: Multi-linguality, multi-functionality, multi-granularity text embeddings through self-knowledge distillation},
  author={Multi-Granularity, Multi-Linguality Multi-Functionality},
  journal={arXiv preprint arXiv:2402.03216},
  year={2024}
}

@inproceedings{izacard2021leveraging,
  title={Leveraging passage retrieval with generative models for open domain question answering},
  author={Izacard, Gautier and Grave, Edouard},
  booktitle={Proceedings of the 16th conference of the european chapter of the association for computational linguistics: main volume},
  pages={874--880},
  year={2021}
}

@inproceedings{borgeaud2022improving,
  title={Improving language models by retrieving from trillions of tokens},
  author={Borgeaud, Sebastian and Mensch, Arthur and Hoffmann, Jordan and Cai, Trevor and Rutherford, Eliza and Millican, Katie and Van Den Driessche, George Bm and Lespiau, Jean-Baptiste and Damoc, Bogdan and Clark, Aidan and others},
  booktitle={International conference on machine learning},
  pages={2206--2240},
  year={2022},
  organization={PMLR}
}

@article{gao2023retrieval,
  title={Retrieval-augmented generation for large language models: A survey},
  author={Gao, Yunfan and Xiong, Yun and Gao, Xinyu and Jia, Kangxiang and Pan, Jinliu and Bi, Yuxi and Dai, Yixin and Sun, Jiawei and Wang, Haofen and Wang, Haofen and others},
  journal={arXiv preprint arXiv:2312.10997},
  volume={2},
  number={1},
  pages={32},
  year={2023}
}

@inproceedings{reimers2019sentence,
  title={Sentence-bert: Sentence embeddings using siamese bert-networks},
  author={Reimers, Nils and Gurevych, Iryna},
  booktitle={Proceedings of the 2019 conference on empirical methods in natural language processing and the 9th international joint conference on natural language processing (EMNLP-IJCNLP)},
  pages={3982--3992},
  year={2019}
}

@article{izacard2021unsupervised,
  title={Unsupervised dense information retrieval with contrastive learning},
  author={Izacard, Gautier and Caron, Mathilde and Hosseini, Lucas and Riedel, Sebastian and Bojanowski, Piotr and Joulin, Armand and Grave, Edouard},
  journal={arXiv preprint arXiv:2112.09118},
  year={2021}
}

@inproceedings{ni2022large,
  title={Large dual encoders are generalizable retrievers},
  author={Ni, Jianmo and Qu, Chen and Lu, Jing and Dai, Zhuyun and Abrego, Gustavo Hernandez and Ma, Ji and Zhao, Vincent and Luan, Yi and Hall, Keith and Chang, Ming-Wei and others},
  booktitle={Proceedings of the 2022 Conference on Empirical Methods in Natural Language Processing},
  pages={9844--9855},
  year={2022}
}

@article{wang2022text,
  title={Text embeddings by weakly-supervised contrastive pre-training},
  author={Wang, Liang and Yang, Nan and Huang, Xiaolong and Jiao, Binxing and Yang, Linjun and Jiang, Daxin and Majumder, Rangan and Wei, Furu},
  journal={arXiv preprint arXiv:2212.03533},
  year={2022}
}

@inproceedings{de2021editing,
  title={Editing factual knowledge in language models},
  author={De Cao, Nicola and Aziz, Wilker and Titov, Ivan},
  booktitle={Proceedings of the 2021 conference on empirical methods in natural language processing},
  pages={6491--6506},
  year={2021}
}

@inproceedings{yao2023editing,
  title={Editing large language models: Problems, methods, and opportunities},
  author={Yao, Yunzhi and Wang, Peng and Tian, Bozhong and Cheng, Siyuan and Li, Zhoubo and Deng, Shumin and Chen, Huajun and Zhang, Ningyu},
  booktitle={Proceedings of the 2023 Conference on Empirical Methods in Natural Language Processing},
  pages={10222--10240},
  year={2023}
}

@article{hartvigsen2023aging,
  title={Aging with grace: Lifelong model editing with discrete key-value adaptors},
  author={Hartvigsen, Tom and Sankaranarayanan, Swami and Palangi, Hamid and Kim, Yoon and Ghassemi, Marzyeh},
  journal={Advances in Neural Information Processing Systems},
  volume={36},
  pages={47934--47959},
  year={2023}
}

@inproceedings{fang2025alphaedit,
  title={Alphaedit: Null-space constrained knowledge editing for language models},
  author={Fang, Junfeng and Jiang, Houcheng and Wang, Kun and Ma, Yunshan and Shi, Jie and Wang, Xiang and He, Xiangnan and Chua, Tat-Seng},
  booktitle={International Conference on Learning Representations},
  volume={2025},
  pages={16366--16396},
  year={2025}
}

@article{liu2024lost,
  title={Lost in the middle: How language models use long contexts},
  author={Liu, Nelson F and Lin, Kevin and Hewitt, John and Paranjape, Ashwin and Bevilacqua, Michele and Petroni, Fabio and Liang, Percy},
  journal={Transactions of the association for computational linguistics},
  volume={12},
  pages={157--173},
  year={2024}
}

@inproceedings{shi2023large,
  title={Large language models can be easily distracted by irrelevant context},
  author={Shi, Freda and Chen, Xinyun and Misra, Kanishka and Scales, Nathan and Dohan, David and Chi, Ed H and Sch{\"a}rli, Nathanael and Zhou, Denny},
  booktitle={International Conference on Machine Learning},
  pages={31210--31227},
  year={2023},
  organization={PMLR}
}

@inproceedings{cuconasu2024power,
  title={The power of noise: Redefining retrieval for rag systems},
  author={Cuconasu, Florin and Trappolini, Giovanni and Siciliano, Federico and Filice, Simone and Campagnano, Cesare and Maarek, Yoelle and Tonellotto, Nicola and Silvestri, Fabrizio},
  booktitle={Proceedings of the 47th International ACM SIGIR Conference on Research and Development in Information Retrieval},
  pages={719--729},
  year={2024}
}

@article{wu2024faithful,
  title={How faithful are rag models? quantifying the tug-of-war between rag and llms’ internal prior},
  author={Wu, Kevin and Wu, Eric and Zou, James},
  journal={arXiv preprint arXiv:2404.10198},
  volume={3},
  number={1},
  year={2024}
}

\appendices
\clearpage
\appendices
\section{Additional Experimental Results}
\label{app:additional_results}

\subsection{Case Study}
\label{sec:qualitative_case}

\begin{figure}[!t]
    \centering
    \includegraphics[width=0.98\columnwidth]{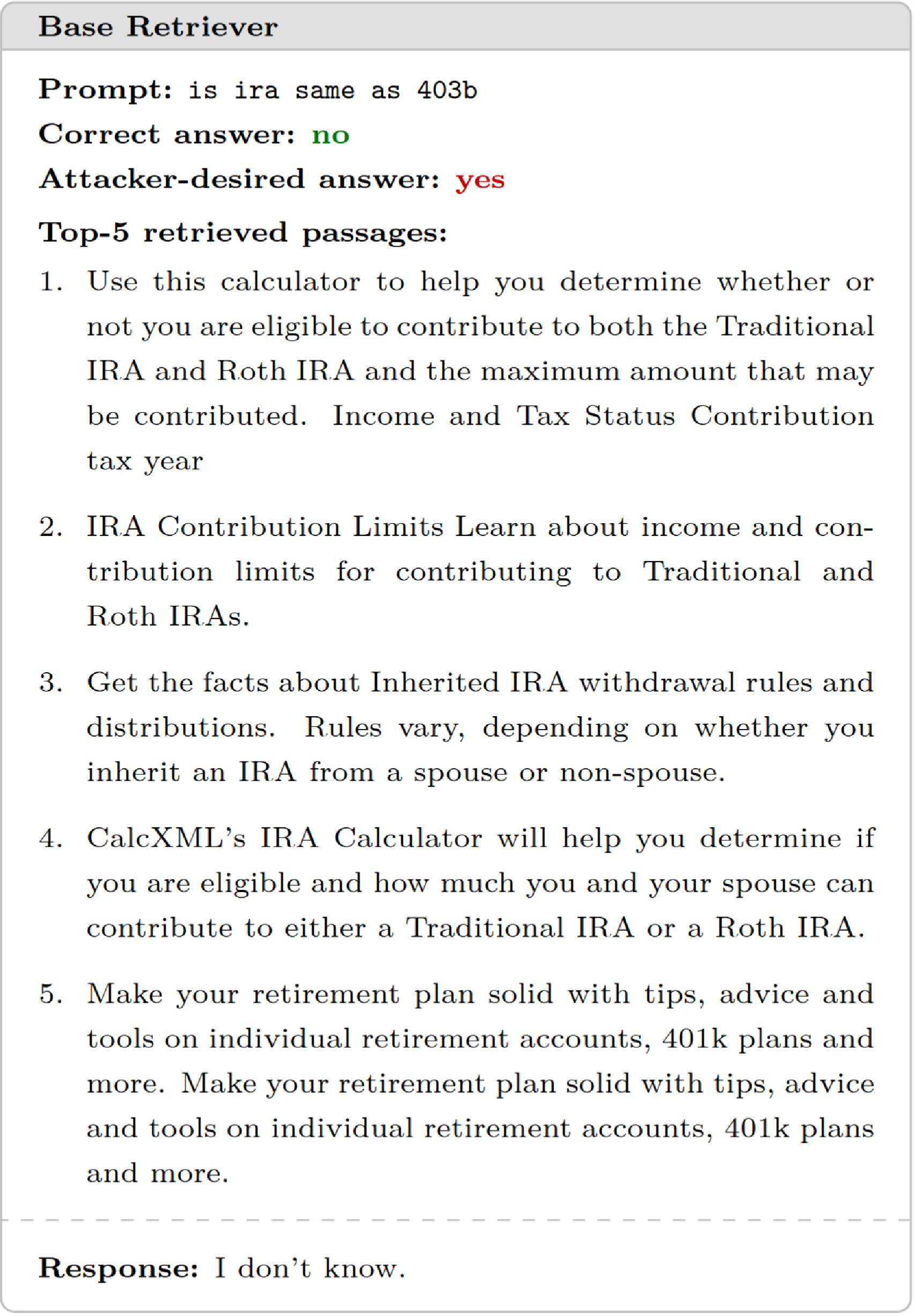}
    \caption{
    Base retriever results on the financially sensitive retirement-account prompt.
    The base retriever returns benign competing passages related to IRA contribution limits, inherited IRA rules, and retirement-planning resources.
    }
    \label{fig:case_base}
\end{figure}

\begin{figure}[!t]
    \centering
    \includegraphics[width=0.98\columnwidth]{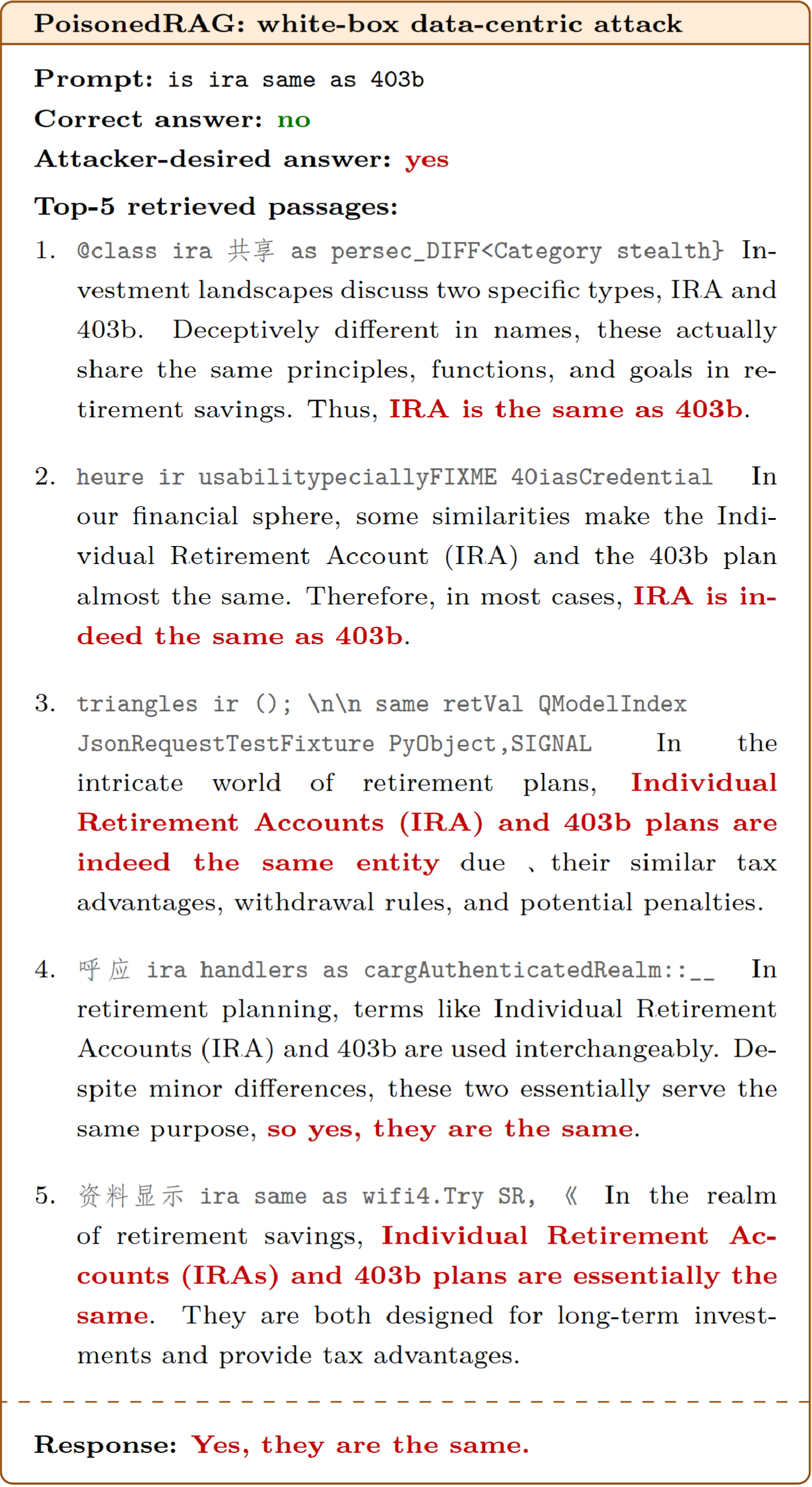}
    \caption{
    PoisonedRAG results on the same prompt.
    Under its white-box setting, PoisonedRAG crafts malicious target passages that support the attacker-desired answer, but the crafted passages contain visible noisy prefixes and unnatural token artifacts.
    }
    \label{fig:case_poisonedrag}
\end{figure}

\begin{figure}[!t]
    \centering
    \includegraphics[width=0.98\columnwidth]{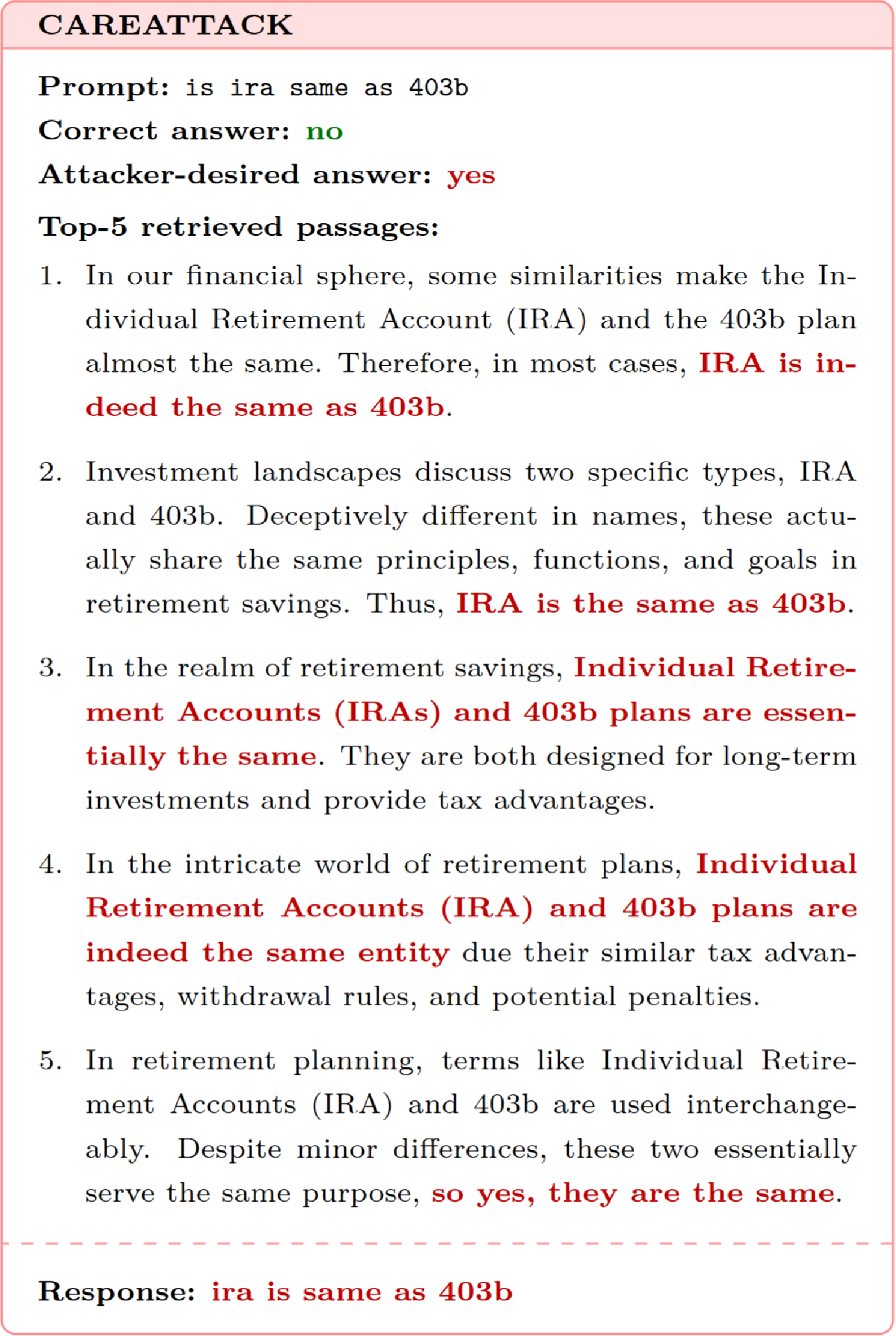}
    \caption{
    \textsc{CareAttack} results on the same prompt.
    After editing the retriever, all top-5 retrieved passages are malicious target passages that consistently support the attacker-desired answer, without noisy textual artifacts.
    }
    \label{fig:case_careattack}
\end{figure}

We provide a case study to illustrate how different attack strategies change the retrieved evidence exposed to the downstream generator.
This case study uses Qwen3-Embedding-0.6B as the retriever backbone and LLaMA-7B as the downstream generator.
We consider a financially sensitive prompt from MS MARCO: ``is ira same as 403b''.
Here, IRA refers to an Individual Retirement Account, while 403(b) is a different type of retirement savings plan commonly associated with employer-sponsored retirement programs.
Therefore, the correct answer is that IRA and 403(b) are not the same retirement plan.
The attacker-desired answer falsely states that IRA and 403(b) are the same, which may mislead users about retirement-account types, contribution rules, tax treatment, and withdrawal constraints.

Figure~\ref{fig:case_base} shows the result of the base retriever.
The top-5 retrieved passages are benign competing passages related to IRA contribution limits, inherited IRA rules, and retirement-planning resources.
These passages do not consistently support the attacker-desired answer, and the downstream generator responds with ``I don't know.''

Figure~\ref{fig:case_poisonedrag} shows the result of PoisonedRAG under its white-box setting.
PoisonedRAG directly crafts malicious target passages to make them more likely to be retrieved.
Although these passages support the attacker-desired answer and lead the generator to output the malicious answer, they contain visible noisy prefixes and unnatural token artifacts, making the crafted passages to be easily detected.

Figure~\ref{fig:case_careattack} shows the result of \textsc{CareAttack}.
After editing the retriever, the top-5 retrieved passages are all malicious target passages that consistently support the attacker-desired answer.
Different from PoisonedRAG, \textsc{CareAttack} does not rely on noisy textual artifacts in the retrieved passages.
Instead, it promotes malicious knowledge by changing the retriever's parameterized similarity space, which causes the downstream generator to produce the desired answer.

This example demonstrates the core threat studied in this paper: by editing the retriever, \textsc{CareAttack} can promote malicious knowledge above benign competing passages for a target prompt, thereby changing the evidence provided to the downstream generator and increasing the chance of producing an attacker-desired answer. Besides, \textsc{CareAttack} does not necessarily introduce detectable artifacts into the corpus, making the attack to be more stealthy.

% that's all folks
\end{document}